\documentclass[11pt]{article}
\usepackage[utf8]{inputenc}
\usepackage[english]{babel}
\usepackage{amsmath}
\usepackage{graphicx}
\usepackage{ae}
\usepackage{fancyvrb}
\usepackage{bbm}
\usepackage{hyperref}
\usepackage{xcolor}
\hypersetup{colorlinks=true,linkcolor=blue,citecolor=black,urlcolor=blue,hypertexnames=true}
\usepackage[margin=1in]{geometry}
\usepackage{float}
\usepackage{graphicx}
\usepackage{chngcntr}
\usepackage{natbib}
\usepackage{amsfonts}
\usepackage{caption}
\usepackage{comment}
\usepackage{multirow}
\usepackage{mathrsfs}

\DefineVerbatimEnvironment{Sinput}{Verbatim}{fontshape=sl}
\DefineVerbatimEnvironment{Soutput}{Verbatim}{}
\DefineVerbatimEnvironment{Scode}{Verbatim}{fontshape=sl}

\DefineVerbatimEnvironment{Code}{Verbatim}{}
\DefineVerbatimEnvironment{CodeInput}{Verbatim}{fontshape=sl}
\DefineVerbatimEnvironment{CodeOutput}{Verbatim}{}
\newenvironment{CodeChunk}{}{}
\setkeys{Gin}{width=0.8\textwidth}
\newlength{\footerskip}
\let\code=\texttt
\let\proglang=\textsf
\newcommand{\pkg}[1]{{\fontseries{m}\fontseries{b}\selectfont #1}}

\title{\pkg{MSTest}: An \proglang{R}-Package for Testing Markov Switching Models}
\date{\today}
\author{Gabriel Rodriguez-Rondon\thanks{This work was supported by the Fonds de recherche sur la société et la culture Doctoral Research Scholarships (B2Z).} \thanks{Mailing address: Department of Economics, McGill University, 855 Sherbrooke St. West Montreal, QC H3A 2T7. e-mail: gabriel.rodriguezrondon@mail.mcgill.ca. Wed page: \url{https://grodriguezrondon.com}} \and  Jean-Marie Dufour\thanks{William Dow Professor of Economics, McGill University, Centre interuniversitaire de recherche en analyse des organisations (CIRANO), and Centre interuniversitaire de recherche en economie quantitative (CIREQ). Mailing address: Department of Economics, McGill University, Leacock Building, Room 414, 855 Sherbrooke Street West, Montreal, Quebec H3A 2T7, Canada. e-mail: jean-marie.dufour@mcgill.ca. Web page: \url{http://www.jeanmariedufour.com}}}

\begin{document}

\maketitle

\begin{abstract}
We present the \proglang{R} package \pkg{MSTest}, which implements hypothesis testing procedures to identify the number of regimes in Markov switching models. These models have wide-ranging applications in economics, finance, and numerous other fields. The \pkg{MSTest} package includes the Monte Carlo likelihood ratio test procedures proposed by \cite{rodrondufour_mcmstest}, the moment-based tests of \cite{dufourluger17}, the parameter stability tests of \cite{chp14}, and the likelihood ratio test of \cite{hansen92}. Additionally, the package enables users to simulate and estimate univariate and multivariate Markov switching and hidden Markov processes, using the expectation-maximization (EM) algorithm or maximum likelihood estimation (MLE). We demonstrate the functionality of the \pkg{MSTest} package through both simulation experiments and an application to U.S. GNP growth data.

\\
%\textbf{JEL Classification:} \\
\end{abstract}
\thispagestyle{empty}
%\newpage
%\tableofcontents
%\thispagestyle{empty}

%%%%%%%%%%%%%%%%%%%%%%%%%%%%%%%%%%%%%%%%%%%%%%%%%%%%%%%%%%%%%%%%%%%%%%%%%%%%%%%%%
%%                           INTRODUCTION                                      %% 
%%%%%%%%%%%%%%%%%%%%%%%%%%%%%%%%%%%%%%%%%%%%%%%%%%%%%%%%%%%%%%%%%%%%%%%%%%%%%%%%%
\newpage
\setcounter{page}{1}

\section[Introduction]{Introduction} \label{sec:intro}
Markov switching models were first introduced by \cite{goldfeld1973markov}, but they were later popularized and became an active area of research in economics after \cite{hamilton89} proposed modeling the first difference of U.S. GNP as a nonlinear stationary process rather than a linear stationary process, as was typically done. The nonlinearity here arises from discrete shifts in the process.

These models have now been considered in various macroeconomic and financial applications. For example, they have been used in the identification of business cycles to provide probabilistic statements about the state of the economy (see \cite{chauvet1998econometric}; \cite{diebold1996measuring}; \cite{kimnel1999}; \cite{chauvet2006dating}; \cite{quqin2021}), in modeling stock market volatility with Markov switching ARCH, GARCH, and Stochastic Volatility models (see \cite{hamilton1994}; \cite{gray1996}; \cite{klaassen2002improving}; \cite{haas2004new}; \cite{pelletier2006regime}; \cite{so1998stochastic}), in modeling interest rate dynamics (see \cite{cai1994markov}; \cite{garper1996}), in considering state-dependent impulse response functions (see \cite{sims2006were}; \cite{caggiano2017}), in the identification of structural shocks in SVAR models (see \cite{lanne2010structural}; \cite{herwartz2014structural}; \cite{lutkepohl2021testing}), and more recently in improving measures of core inflation with multiple inflation regimes (see \cite{rodron_inf_2024}). \cite{hamilton2016} provides a detailed survey of regime switching in macroeconomics.

Outside of macroeconomic and financial applications, these models have also been applied in climate change research (see \cite{golosov2014optimal}; \cite{dietz2015endogenous}), environmental and energy economics (see \cite{cevik2021renewable}; \cite{charfeddine2017impact}), industrial organization (see \cite{aguirregabiria2007sequential}; \cite{sweeting2013dynamic}), and health economics (see \cite{hernandez2016switching}; \cite{anser2021impact}). Additionally, there is a related model—the Hidden Markov model—which has various applications in computational molecular biology (see \cite{krogh1994hidden}; \cite{baldi1994hidden}), handwriting and speech recognition (see \cite{rabiner1986introduction}; \cite{nag1986script}; \cite{RabJuanFundamentals}; \cite{jelinek1997statistical}), computer vision and pattern recognition (see \cite{bunke2001hidden}), and other machine learning applications.

Given their empirical relevance, it is important to determine the number of regimes needed to properly capture the nonlinearities present in the data, as this is not determined endogenously when estimating Markov switching models. However, the asymptotic results of conventional hypothesis testing procedures do not apply in this setting because the regularity conditions required for such results are violated. Consequently, alternative hypothesis testing procedures have been proposed in the literature.

Notable contributions to testing the null hypothesis of a linear model against a model with two regimes include \cite{hansen92}, \cite{hansen96}, \cite{garcia98}, \cite{chowhite07}, \cite{marmer2008}, \cite{chp14}, \cite{kso2014modqlr}, \cite{dufourluger17}, and \cite{quzhuo2021likelihood}. Testing the null of an $M$-regime model against the alternative of an $M+m$-regime model for $M \geq 1$ and $m = 1$ has been considered by \cite{kasshi2018}, where the authors show that the parametric bootstrap test can be asymptotically valid when imposing certain restrictions on the parameter space and specifically considering univariate models with fixed or predetermined regressors. \cite{quzhuo2021likelihood} present similar results regarding the parametric bootstrap for $M = m = 1$ but for a broader set of univariate models, however, still containing the parameter space. More recently, \cite{rodrondufour_mcmstest} propose Monte Carlo likelihood ratio tests that handle cases where both $M \geq 1$ and $m \geq 1$ and also consider multivariate settings, which had not been addressed previously. Their test procedures are the most general procedure available and deal transparently with issues related to violations of regularity conditions. Importantly, they do not require parametric restrictions, normality of errors, or stationary processes, as the existence of an asymptotic distribution is unnecessary. This makes the tests applicable in more cases than the parametric bootstrap and allows for use in settings where the asymptotic validity of the parametric bootstrap has not been established. The maximized Monte Carlo version of their test even controls the test size in finite samples, which is particularly relevant for many macroeconomic applications using quarterly data. Additionally, this version of their test is robust to identification problems, which are common when working with Markov switching models.

Testing the number of regimes that a Markov switching model should include is an important step when deciding on the model's specification. However, performing these tests is not necessarily trivial, and conducting more than one test can quickly become tedious. As a result, we introduce \pkg{MSTest}, an \proglang{R} package that can be used to test the null hypothesis of $M$ regimes against the alternative hypothesis of $M+m$ regimes for both univariate and multivariate models. The purpose of this \proglang{R} package is to enable econometricians to determine the number of regimes that a model should include for a given process by making hypothesis testing procedures readily available to a general audience. It aims to facilitate the comparison of different test procedures and the determination of the number of regimes in a model for economic research. The package also allows users to simulate and estimate univariate and multivariate Markov switching models, as well as hidden Markov models. Estimation is provided through the use of the expectation maximization (EM) algorithm or maximum likelihood estimation (MLE). \pkg{MSTest} utilizes Rcpp (\cite{eddetal2018}) and RcppArmadillo (\cite{eddsan2024}) for computational efficiency. This is especially important given the computational burden of testing in the presence of nuisance parameters, as is the case here. The \pkg{MSTest} package includes the methodologies presented in \cite{rodrondufour_mcmstest}, \cite{dufourluger17}, \cite{chp14}, and \cite{hansen92}. The parametric bootstrap discussed by \cite{quzhuo2021likelihood} and \cite{kasshi2018} is not explicitly provided but can be performed using specific settings while employing the local Monte Carlo likelihood ratio test of \cite{rodrondufour_mcmstest}. These testing procedures, along with other notable contributions previously mentioned, are discussed in more detail in Section (\ref{sec:testing}) of this paper.

In Section \ref{sec:models}, we describe Markov switching and hidden Markov models for which the hypothesis test are implemented. In Section \ref{sec:testing}, we describe some theoretical aspects of the hypothesis testing procedures the \pkg{MSTest} package offers. That is, we discuss the identification difficulties in more detail and the methodologies developed in each framework to test the hypothesis of interest. Section \ref{sec:software} describes the package in more detail and the functions available to test for the number of regimes. This section is meant to compliment the \pkg{MSTest} documentation available through CRAN, by giving a general overview of available functions and can also serve as a short manual for the \pkg{MSTest} package. Section \ref{sec:empiricalex} includes an empirical example where we apply the test procedures included in \pkg{MSTest} to the U.S. GNP data of \cite{hamilton89}, the extended data used in \cite{chp14} and \cite{dufourluger17} and a further extension of the data ranging from 1951Q2 to 2024Q2. We compare the different models and provide tables with: test statistics, critical values and p-values. Finally we provide brief concluding remarks in section \ref{sec:conclusion}.

%% -- Markov switching Models --------------------------------------------------
\section[Markov switching models]{Markov switching models} \label{sec:models}

The \pkg{MSTest} package examines Markov switching models where only the mean and variance are governed by the Markov process  $S_t$ and so here, we describe such models. We also consider a specific case of the Markov switching model—the Hidden Markov model—in which no autoregressive coefficients are included as explanatory variables. In both cases, other exogenous explanatory variables may be included.
 
% ----------------------------------------- %
% ----- Markov process
% ----------------------------------------- %
\subsection{First-order Markov process}
We begin by describing the first-order Markov process $S_t$ that governs the changes in the parameters of the Markov switching model. We assume that the process $S_t$ is unobserved and evolves according to a first-order ergodic Markov chain with a ($M \times M$) transition probability matrix given by 
\begin{align*}
    \textbf{P} & = \begin{bmatrix}
            p_{11} & \dots & p_{M1}\\
            \vdots & \ddots & \vdots \\
            p_{1M} & \dots & p_{MM}
    \end{bmatrix}
\end{align*}
where for example $p_{ij} = P(S_t = j | S_{t-1} = i)$ is the probability of state $i$ being followed by state $j$ and $M$ is the total number of regimes. Specifically, if we consider $M$ regimes, the process takes integer values $S_t=\{1, \dots, M\}$. Additionally, the columns of the transition matrix $\textbf{P}$  must sum to one in order to have a well defined transition matrix (i.e., $\sum^M_{j=1} p_{ij} = 1$).
 
Considering the example in \cite{hamilton1994} where the Markov process has only two regimes, we only need a $(2 \times 2)$ transition matrix to summarize the transition probabilities $\textbf{P}$ as follows:
\begin{align*}
    \textbf{P} & = \begin{bmatrix}
            p_{11} & p_{21}\\
            p_{12} & p_{22}
    \end{bmatrix}
\end{align*}
We can also obtain the ergodic probabilities, $\pi = (\pi_1, \pi_2)'$, which are given by
\begin{align*}
    \pi_1 & = \frac{1-p_{22}}{2-p_{11}-p_{22}} & \pi_2 & = 1-\pi_1 %\label{ergodic_prob}
\end{align*}
in a setting with two regimes. More generally, for any number of $M$ regimes we could use  
\begin{align*}
    \pmb{\pi} = (\mathbf{A}'\mathbf{A})^{-1}\mathbf{A}'\mathbf{e}_{N+1} \hspace{0.25cm}  \&  \hspace{0.25cm} \mathbf{A} = \begin{bmatrix}
        \mathbf{I}_M-\textbf{P}\\
        \pmb{1}'
        \end{bmatrix}
\end{align*}
where $\mathbf{e}_{M+1}$ is the $(M+1)$th column of $\mathbf{I}_{M+1}$. These ergodic probabilities tell us on average, in the long-run, the proportion of time the process $S_t$ spends in each regime. 
% ----------------------------------------- %
% ----- Markov switching autoregressive model
% ----------------------------------------- %
\subsection{Markov switching autoregressive models}
The Markov switching model considered in the \pkg{MSTest} package can be expressed as
\begin{align}
    y_t & = \mu_{S_t} + \sum^{p}_{k=1} \phi_{k} (y_{t-1} - \mu_{S_{t-1}}) +  Z_t\beta_z + \sigma_{S_t}\epsilon_t \label{msm_uni}
\end{align}
where, in a univariate setting, $y_t$ is a scalar, $Z_t$ is a $(1 \times q_{z})$ vector of exogenous variables whose coefficients do not depend on the latent Markov process $S_t$, and $\epsilon_t$ represents the error process, which, for example, may be distributed as a $\mathcal{N}(0,1)$. The error term is multiplied by the standard deviation $\sigma_{S_t}$, which may either depend on the Markov process or remain constant throughout (i.e., $\sigma$).

This Markov switching autoregressive model is labeled as ``\code{MSARmdl}'' in the \pkg{MSTest} package when $Z_t$ is excluded, or as ``\code{MSARXmdl}'' when exogenous regressors, $Z_t$, are included. These are the versions most commonly used in various economic and financial applications, as well as other time series-related applications. Other error distributions, such as a Student-t distribution, may also be considered in future versions of the package. Currently, the \pkg{MSTest} package only considers Markov switching autoregressive models with normally distributed errors when simulating the processes, so we focus on this setup in this paper.

Continuing with the example where a Markov switching model given by equation (\ref{msm_uni}) has $M=2$ regimes, such that $S_t = \{1, 2\}$, the sample log likelihood conditional on the first $p$ observations of $y_{t}$ is given by 
\begin{align}
    L_{T} (\theta) = \mathrm{log} f(y_{1}^{T}|y^{0}_{-p+1};\theta) = \sum^{T}_{t=1} \mathrm{log} f(y_{t}|\mathscr{Y}_{t-1};\theta) \label{arms_loglik}
\end{align}
where $\mathscr{Y}_{t-1} = \sigma$-field$\{\dots,Z_{t-1},y_{t-2}, Z_{t}, y_{t-1}\}$ and $\theta = (\mu_{1}, \mu_{2}, \beta, \sigma_{1}, 
\sigma_{2}, vec(\textbf{P}))$ and $vec(\cdot)$ is the vectorization operator which stacks the columns of a matrix to form a column vector. Here, 
\begin{align}
    f(y_{t}|\mathscr{Y}_{t-1};\theta) = \sum^2_{s_{t}=1}\sum^2_{s_{t-1}=1}\dots \sum^2_{s_{t-p}=1} f(y_{t},S_{t}=s_{t},S_{t-1}=s_{t-1},\dots,S_{t-p}=s_{t-p}|\mathscr{Y}_{t-1};\theta) \label{arms_loglik_f}
\end{align}
and more specifically 
\begin{equation}
    \begin{aligned}
    f(y_{t},S_{t}=s_{t},\dots,S_{t-p}= s_{t-p}|\mathscr{Y}_{t-1};\theta) = & \frac{\text{Pr}(S^{*}_{t}=s^{*}_{t}|\mathscr{Y}_{t-1};\theta)}{\sqrt{2\pi\sigma_{s_{t}}^2}} \\
    & \times \text{exp}\left\{\frac{-[(y_{t} - \mu_{s_t}) - \sum^{p}_{k=1} \phi_{k} (y_{t-1} - \mu_{s_{t-1}}) -  Z_t\beta_z ]^2}{2\sigma_{s_{t}}^2}\right\}
    \end{aligned}
\label{normal_msm_pdf}
\end{equation}
where we set 
\begin{align*}
    S^{*}_{t} = s^{*}_{t} \hspace{0.25cm} \text{if} \hspace{0.25cm} S_{t} = s_{t}, S_{t-1} = s_{t-1}, \dots, S_{t-p} = s_{t-p}   
\end{align*}
and $\text{Pr}(S^{*}_{t}=s^{*}_{t}|\mathscr{Y}_{t-1};\theta)$ is the probability that this occurs. Note that as in, \cite{rodrondufour_mcmstest}, here we denote the latent variable that determines the regimes at time $t$ as $S_{t}$ and let $s_{t}$ denote the (observed) realization of $S_{t}$.
% ----------------------------------------- %
% ----- Markov switching VAR model
% ----------------------------------------- %
\subsection{Markov switching VAR model}
\cite{krolzig1997markov} generalized the the univariate Markov switching autoregressive model to the multivariate setting and hence introduced the Markov switching Vector Autoregressive (MS-VAR) model. The MS-VAR model considered in the \pkg{MSTest} package can be expressed as
\begin{align}
    \pmb{y}_t & = \pmb{\mu}_{S_t} + \pmb{\Phi}_1(\pmb{y}_{t-1} - \pmb{\mu}_{S_{t-1}}) + \cdots + \pmb{\Phi}_p(\pmb{y}_{t-p} - \pmb{\mu}_{S_{t-p}}) + Z_{t}\pmb{\beta} + \pmb{\Sigma}^{1/2}_{S_t}\pmb{\epsilon}_t \label{msm_multi}
\end{align}
where $\pmb{y}_t = [y_{1,t}, \dots, y_{q,t}]'$, $\pmb{\mu}_{S_t} = [\mu_{1,S_t}, \dots, \mu_{q,S_t}]'$, $\pmb{\epsilon}_t = [\epsilon_{1,t}, \dots, \epsilon_{q,t}]'$, $\pmb{\Phi}_k$ is a ($q \times q$) matrix containing the autoregressive parameters at lag $k$, $\pmb{\beta}$ is now a ($q_z \times q$) matrix, and $\pmb{\Sigma}_{S_t}=\pmb{\Sigma}^{1/2}_{S_t}(\pmb{\Sigma}^{1/2}_{S_t})'$ is the ($q \times q$) regime dependent covariance matrix. As with the univariate setting, the \pkg{MSTest} package also includes a version without exogenous regressors $Z_{t}$, labeled ``\code{MSVARmdl}'', and a version which allows the inclusion of exogenous regressors, labeled ``\code{MSVARXmdl}''. More sophisticated versions of the MS-VAR model, as well as their likelihood functions, are described in \cite{krolzig1997markov} and we direct the interested reader to consider this reference to learn more about these models and their components.
% ----------------------------------------- %
% ----- Hidden Markov model
% ----------------------------------------- %
\subsection{Hidden Markov model}
Hidden Markov models (HMM) can be shown to be a special case of the more general Markov switching model we defined above. Specifically, Hidden Markov models don't necessarily have to be applied to time series data and for this reason they typically do not include lags of the endogenous variable, $y_{t}$. 

For example, we can recover a hidden Markov from (\ref{msm_multi}) by simply excluding lags of $y_t$ as explanatory variables giving 
\begin{align}
    \pmb{y}_t & = \pmb{\mu}_{S_t} + Z_{t}\pmb{\beta} + \sigma_{S_t}\epsilon_t \label{hmm}
\end{align}
When $q=1$, we recover a univariate Hodden Markov model from (\ref{msm_uni}). This version and its multivariate counterpart are the HMMs considered in the package \pkg{MSTest} and are labeled as ``\code{HMmdl}''. 

As described by \cite{an2013identifiability}, the dependence on past observations allows for more general interactions between $y_t$ and $S_t$ which can be used to model more complicated causal links between economic or financial variables of interest. Including past observations is a very common practice in economic time series applications as a way to control for stochastic trends, which may explain why Markov-switching models are more popular than the basic HMM in this literature. 
% ----------------------------------------- %
% ----- Model Estimtion
% ----------------------------------------- %
\subsection{Model estimation}
Typically, Markov switching models are estimated using the Expectation-Maximization (EM) algorithm (see \cite{dempster_maximum_1977}), Bayesian methods, or with the Kalman filter when employing the state-space representation of the model. In very simple cases, Markov switching models can be estimated using Maximum Likelihood Estimation (MLE). However, since the Markov process $S_{t}$ is latent and, more importantly, because the likelihood function may exhibit several modes of equal height along with other unusual features that complicate MLE estimation, this approach is less commonly used.

The \proglang{R} package \pkg{MSTest} enables estimation of the models described above via the EM algorithm by setting ``\code{control = list(method = `EM')}'' or via MLE by setting ``\code{control = list(method = `MLE')}'' in the estimation functions, which are detailed below in Section \ref{sec:software}. In practice, empirical estimates can sometimes be improved by using the EM algorithm results as initial values in a Newton-type optimization algorithm. This two-step estimation procedure is used to obtain results presented in the empirical section of \cite{rodrondufour_mcmstest}, and in other related works.

We omit a detailed explanation of the EM algorithm and MLE, as our focus is on describing the \pkg{MSTest} package. For the interested reader, the estimation of a Markov switching model via the EM algorithm and MLE is described in detail in \cite{hamilton1990} and \cite{hamilton1994}, and for Markov-switching VAR models, in \cite{krolzig1997markov}.

%% -- Hypothesis testing procedures -------------------------------------------
\section[Hypothesis testing for number of regimes]{Hypothesis testing for number of regimes} \label{sec:testing}
When estimating a Markov switching model, it is essential to determine the number of regimes to be estimated, as this is not decided endogenously during the estimation process. However, it is well understood in the literature that when considering testing for the number of regimes in a Markov switching model, conventional hypothesis testing procedures are no longer valid as the typical regularity conditions needed for asymptotic validity are not met. To address these challenges, various studies have proposed alternative methods that yield valid testing procedures. In this section, we describe some key procedures, focusing on those included in the \proglang{R} package \pkg{MSTest}. As mentioned in the introduction, the purpose of \pkg{MSTest} is to make the most useful of these procedures accessible to a general audience. Here, we explain how these procedures fit within the existing literature on hypothesis testing for Markov switching models and briefly review how these procedures circumvent these violations of regularity conditions.

In general, the hypothesis test of interest is 
\begin{align*}
    H_0: M & = M_0  \\
    H_1: M & = M_0 + m
\end{align*}
where both \(M_0, m \geq 1\). However, most available test procedures can only address the case when $M_0=m=1$ and so 
\begin{align*}
    H_0: M & = 1  \\
    H_1: M & = 2
\end{align*}
In this case, a linear model (one regime) is being considered under the null hypothesis and is being compared against a Markov switching model with two regimes under the alternative hypothesis. Currently, the most general procedure able to deal with settings where \(M_0, m \geq 1\) are the Monte Carlo likelihood ratio tests described in \cite{rodrondufour_mcmstest} but this flexibility comes at the cost of, relatively speaking, being computationally intensive. For this reason \pkg{MSTest} also includes other procedures such as the moment-based tests of \cite{dufourluger17} and the parameter stability test of \cite{chp14}, which are computationally efficient and useful when only considering the more simple case of $M_0=m=1$. Moreover, the package includes the standardized likelihood ratio test of \cite{hansen92}, which is a significant contribution and has frequently served as a benchmark for evaluating other test procedures.

% ----------------------------------------- %
% ----- LRT
% ----------------------------------------- %
\subsection{Likelihood ratio tests}
\cite{hansen92} was the first to propose a testing procedure for Markov switching models when $M_0=m=1$, so we begin with a review of this procedure, which is available in \pkg{MSTest}. The author provides a thorough description of the problems that plague the likelihood ratio approach for testing the number of regimes in a Markov switching model. First, it is typically assumed that the likelihood function is locally quadratic in the region where the null hypothesis and the globally optimal estimated parameters are found. However, as the author notes, since some parameters are not identified under the null, this region is likely flat with respect to those unidentified parameters rather than quadratic. Unidentified nuisance parameters under the null hypothesis are issues that have been considered in \cite{davies1977}, \cite{davies87}, \cite{andrewsploberger94}, and \cite{dufour2006}. Second, it is commonly assumed that the score is positive; however, as described, it can be identically $0$ under the restricted maximum likelihood estimator (MLE) of a linear model (the null hypothesis). Third, some parameters, such as the transition probabilities, can take values of $0$ and $1$, which leads to the parameter boundary problem discussed in \cite{andrews1999} and \cite{andrews2001}. Additionally, the likelihood surface can have multiple local optima, meaning the null hypothesis may not lie in the same region as the global optimum.

\cite{hansen92} introduces a new approach that does not require the conventional assumptions associated with likelihood ratio tests. Instead, the authors model the likelihood function as an empirical process of the unknown parameters. They utilize empirical process theory to establish a bound for the asymptotic distribution of the standardized likelihood ratio test. \cite{hansen92} formulates the hypothesis as follows:

\begin{equation}
    \begin{aligned}
        H_0: \alpha = \alpha_0 \\
        H_1: \alpha \neq \alpha_0
    \end{aligned}
\end{equation}
\noindent
where $\alpha_0$ represents the parameter values under the null and $\alpha$ the parameter values under the alternative. They begin by decomposing the likelihood ratio as such: 
\begin{equation}
\begin{aligned}
    LR_n(\alpha)& = L_n(\alpha) - L_n(\alpha_0) \\
                & = \Sigma_{i=0}^n [l_i(\alpha) - l_i(\alpha_0)] \\
                & = R_n(\alpha)  + Q_n(\alpha)
\end{aligned}
\end{equation}
\noindent
where $R_n(\alpha) = E[LR_n(\alpha)]$ is the expectation of the likelihood ratio function and $Q_n(\alpha) = \Sigma_{i=1}^n q_i(\alpha) = [l_i(\alpha) - l_i(\alpha_0)] - E[l_i(\alpha) - l_i(\alpha_0)]$ is the deviations from the mean. Fluctuations in Q play an important role in identifying an optimum as:
\begin{equation}
\begin{aligned}
    \frac{1}{\sqrt{n}} LR_n (\alpha)& = \frac{1}{\sqrt{n}} R_n (\alpha) + \frac{1}{\sqrt{n}} Q_n (\alpha) \\
                                    & =\frac{1}{\sqrt{n}} R_n (\alpha) +Q (\alpha) + o_p(1)
\end{aligned}
\end{equation}
\noindent
and by using the fact that \(R_n(\alpha) \leq 0 \) for all \(\alpha\) when the null hypothesis is true, we can see that \(\frac{1}{\sqrt{n}} LR_n (\alpha) \leq \frac{1}{\sqrt{n}} Q_n (\alpha)\) and so it follows that,
\begin{equation}
    P[\frac{1}{\sqrt{n}} LR_n \geq x] \leq P[\sup_{\alpha}\frac{1}{\sqrt{n}} Q_n (\alpha) \geq x] \xrightarrow{} P[\sup_{\alpha} Q(\alpha) \geq x]
\end{equation}
Thus, the distribution of the empirical process Q can provide a bound for the asymptotic distribution of the LR statistic. The test statistic is further standardized:
\begin{equation}
    LR^*_n = \sup_{\alpha} LR^*_n(\alpha)
    \label{hansen_stdLRT}
\end{equation}
\noindent 
where
\begin{equation}
    LR^*_n(\alpha) = \frac{LR_n(\alpha)}{V_n(\alpha)^{1/2}} 
\end{equation}

As suggested by equation (\ref{hansen_stdLRT}), they resolve the issue of nuisance parameters by evaluating the standardized LR statistic for different parameter values $\alpha$. Specifically, they evaluate the test statistic over a grid of different parameter values and optimize with respect to those nuisance parameter values. To be clear, \cite{hansen92} sets $\alpha = (\mu_2, \sigma_2, p_{11},p_{22})$ as the vector of nuisance parameters, which includes the second state parameters and transition parameters. The first regime parameters $\theta=(\mu_1,\sigma_1,\phi_1,\dots,\phi_p)$ are fully identified. They further split $\alpha$ into $\beta =(\mu_2, \sigma_2)$ and $\gamma = (p_{11},p_{22})$, where $\beta$ is treated as a parameter of interest and set to be identical to $\mu_1$ and $\sigma_1$ under the null. However, the process $Q$ may also be serially correlated for some values of $\alpha$, and so \cite{hansen96} adds a correction that should be used for calculating the asymptotic distribution of the test statistic, which is also included in the implementation of this test procedure in \pkg{MSTest}.

There are two main drawbacks to consider when using this likelihood ratio procedure for testing Markov switching models. The first drawback is that this test only provides a bound for the standardized LR statistic, which can be conservative. Therefore, it is important to remember that the critical values provided by this test in the \pkg{MSTest} package are not necessarily those of the standardized LR statistic but rather of the process $Q$, which provides a bound for the standardized LR statistic. The second drawback is that it involves optimizing the value of the nuisance parameters through a grid search. Although this process is manageable when analyzing only a few variables, as it requires switching between regimes in addition to the transition parameters $p_{11}$ and $p_{22}$, it can quickly become computationally intensive for models that consider more parameters as switching between regimes. Despite some of these drawbacks, we include this test in \pkg{MSTest} because its properties are well understood and it has often been used as a benchmark for comparison in the literature.

%%%%%%%%% Garcia
\cite{garcia98} also reviewed the problem of testing the number of regimes in a Markov switching models using a likelihood ratio approach. \cite{garcia98} builds on \cite{hansen92}'s approach but differs in that they only treat $\gamma$ as nuisance parameters over which we must optimize. This change simplifies some of the computational burden. The $\beta$ parameter remains a parameter of interest but is incorporated into $\theta$, the identified parameters. Although this is a significant contribution and the authors provide valuable insights, this test is not included in this version of \pkg{MSTest}. One reason is that the author assumes that the LR test can be expressed as the supremum of a chi-square functional asymptotically under the null hypothesis, a claim that \cite{hansen96inference} and \cite{andrewsploberger94} suggest cannot be made for Markov switching models. However, it may still be included in future versions of \pkg{MSTest} for completeness.

%%%%%%%%% Cho and White
\cite{chowhite07} also considers hypothesis testing when the null hypothesis is a linear model and the alternative is a Markov switching model with two regimes. They address a difficulty not covered by \cite{hansen92} or \cite{garcia98}, specifically the case where parameters lie on the boundary of the parameter space. The authors classify the null hypothesis into two mutually exclusive subsets: one where $p \in (0,1)$ and another where $p = 0$ or $p = 1$, where $p$ is a transition probability (e.g., $p_{11}$). The second case presents the boundary parameter problem. Building on the work of \cite{andrews1999} and \cite{andrews2001}, the authors develop a QLR test statistic that accounts for this boundary issue. Additionally, \cite{chowhite07} employs a normal mixture model framework to formulate the QLR test within this context. They argue that this approach allows them to disregard certain time series dependence properties implied by the Markov regime-switching process. The authors describe the QLR test as sensitive to the mixture aspect of the regime-switching process, claiming it delivers a test with strong power under the alternative hypothesis. A significant finding of their work is that the critical values of the asymptotic distribution of the test statistic are influenced by the consideration of the boundary problem. In other words, they demonstrate the importance of addressing the boundary problem when examining the asymptotic distribution of the test statistic. However, \cite{carsteig2012} highlight a critical flaw in this test, arguing that it may overlook the time dependency of the Markov chain but fails to account for the time dependencies present in an autoregressive model when parameters change across regimes. The authors acknowledge this limitation in \cite{chowhi2011}. For this reason, the test proposed by \cite{chowhite07} is not currently included in \pkg{MSTest}, as autoregressive models are particularly relevant in many economic and financial applications.

%%%%%%%%% Qu and Zhuo 2021
\cite{quzhuo2021likelihood} presents a novel characterization of the conditional regime probabilities for a family of likelihood ratio-based tests and establishes the asymptotic distribution for the test statistics. Like \cite{hansen92}, \cite{garcia98}, and \cite{chowhite07}, they derive an approximation of the likelihood ratio as an empirical process, where $\{(p_{11},p_{22}): \epsilon \leq p_{11},p_{22} \leq 1-\epsilon \ \& \ p_{11}+p_{22}\geq 1+\epsilon\}$ and $\epsilon$ is a small constant. They also provide a finite sample refinement to correct some of the over-rejections that can occur in specific cases. As a result, they are able to study the asymptotic null distribution and find that, even though the null hypothesis has only one regime and thus parameters do not switch, the nuisance parameters can affect the limiting distribution, which will depend on which parameters are allowed to switch. Furthermore, \cite{quzhuo2021likelihood} describe how some of these results can explain why specific bootstrap procedures may be inconsistent (e.g., when including weakly exogenous regressors) and why standard information criteria such as the BIC can be sensitive to the hypothesis and the model structure. Although the \proglang{R} package \pkg{MSTest} currently does not include the test procedure proposed by \cite{quzhuo2021likelihood}, future versions will likely incorporate this procedure, as it represents a significant contribution and provides the best approximation to the asymptotic distribution of the likelihood ratio test statistic, making it useful for users interested in these asymptotic results.

%%%%%%%%% Kasahara and Shimotsu
Another likelihood ratio-based test is that of \cite{kasshi2018}. In \cite{kasshi2015}, the authors introduce a re-parameterization and higher-order expansion of the likelihood ratio function. In \cite{kasshi2018}, they apply this technique along with the Difference in Quadratic Mean (DQM) approximation introduced by \cite{liushao2003} to address some of the issues discussed above while estimating and studying the asymptotic distribution of the likelihood ratio test statistic for Markov switching models. Additionally, they do this in a more general setting where $M_0 \geq 1$ but $m=1$ still, allowing them to consider a null hypothesis with more than one regime. This makes their approach more general than the likelihood ratio test procedures discussed so far in that regard. In doing so, like \cite{quzhuo2021likelihood}, they also show that the parametric bootstrap procedure is an asymptotically valid test procedure, but in this case, they only consider the scenario where fixed or predetermined regressors are included in the model while examining Markov switching autoregressive models. The parametric bootstrap procedure is not directly available in \pkg{MSTest}, but it can be implemented using specific options within the Local Monte Carlo likelihood ratio test (LMC-LRT) proposed by \cite{rodrondufour_mcmstest}, which is included. This process is further detailed in section \ref{sec:software}, where we discuss the test procedures available within the \pkg{MSTest} package. Notably, the Monte Carlo procedures of \cite{rodrondufour_mcmstest} are applicable in even more settings (e.g., $m>1$ and more) and so they are discussed next. 

%%%%%%%%% Rodriguez-Rondon and Dufour 
In \cite{rodrondufour_mcmstest}, the authors propose the Maximized Monte Carlo likelihood ratio test (MMC-LRT) and the Local Monte Carlo likelihood ratio test (LMC-LRT), which can be used to compare very general Markov switching models. These procedures represent the most general type of testing methods, as they can be applied to hypothesis testing in its broadest form when both $M_{0}$ and $m$ are greater than or equal to 1. Furthermore, as described in \cite{rodrondufour_mcmstest}, these Monte Carlo likelihood ratio tests can be utilized when the process $y_{t}$ is non-stationary, when the model exhibits non-Gaussian errors, when parameters take values at the boundary, and in multivariate settings, which include the previously discussed Markov switching VAR and multivariate Hidden Markov models. To be more precise, the MMC-LRT and LMC-LRT are the only test procedures available in the literature and in the \proglang{R} package \pkg{MSTest} that can be employed to test multivariate Markov switching models. Additionally, the MMC-LRT procedure is valid in finite samples and is robust to identification issues, features that are empirically relevant when considering macroeconomic applications of Markov switching models, as discussed in \cite{rodrondufour_mcmstest}.
 
Here, we summarize the MMC-LRT and LMC-LRT procedures but readers interested in further details are referred to the more formal description provided in \cite{rodrondufour_mcmstest} and \cite{dufour2006} for even further details on the Monte Carlo techniques used in these procedures. For simplicity of exposition, we use an example where we are interested in a null hypothesis of a linear model (\emph{i.e.}, only $M_{0}=1$ regime) and an alternative hypothesis of $M_{0}+m=2$ regimes and consider an autoregressive model where only the mean and variance are subject to change. Since this is a likelihood ratio-based approach, the log-likelihood values under the null and alternative hypothesis are required. The log-likelihood for the model under the alternative (and under the null hypothesis if $M_{0}>1$) is given by (\ref{arms_loglik}) - (\ref{normal_msm_pdf}): 
\begin{equation}
    L_{T}(\theta _{1})=\mathrm{log}\,f(y_{1}^{T}\,|\,y_{-p+1}^{0};\,\theta_{1})=\sum_{t=1}^{T}\mathrm{log}\,f(y_{t}\,|\,y_{-p+1}^{t-1};\,\theta _{1})
\end{equation}%
where%
\begin{equation}
    \theta _{1}=(\mu _{1},\,\mu _{2},\,\sigma _{1},\,\sigma _{2},\,\phi_{1},\,\ldots \,,\,\phi _{p},\,p_{11},\,p_{22})^{\prime }\in \Omega \,.
\end{equation}%
Here, the subscript of $1$ underscores the fact that $\theta _{1}$ is the parameter vector under the alternative hypothesis. The set $\Omega$ satisfies any theoretical restrictions we may wish to impose on $\theta _{1}$ [such as $\sigma _{1}>0$ and $\sigma _{2}>0$]. On the other hand, the log-likelihood under the null hypothesis ($M_{0}=1$) is given by 
\begin{equation}
    L_{T}^{0}(\theta _{0})=\mathrm{log}\,f(y_{1}^{T}\,|\,y_{-p+1}^{0};\,\theta_{0})=\sum_{t=1}^{T}\mathrm{log}\,f(y_{t}\,|\,y_{-p+1}^{t-1};\,\theta _{0})
\label{ar_loglik}
\end{equation}%
where 
\begin{equation}
    f(y_{t}\,|\,y_{-p+1}^{t-1};\,\theta _{0})=\frac{1}{\sqrt{2\pi \sigma ^{2}}} \mathrm{exp}\left\{ \frac{-[y_{t}-\mu -\sum_{k=1}^{p}\phi _{k}(y_{t-k}-\mu)]^{2}}{2\sigma ^{2}}\right\} \,,
\end{equation}%
\begin{equation}
    \theta _{0}=(\mu ,\,\sigma ^{2},\phi _{1},\,\ldots \,,\,\phi _{p})^{\prime}\in \bar{\Omega}_{0}.
\end{equation}%
Note that $\bar{\Omega}_{0}$ has lower dimension than $\Omega $. The null and alternative hypotheses can be written as:%
\begin{equation}
    H_{0}:\delta _{1}=\delta _{2}=\delta \quad \text{for some unknown $\delta=(\mu $, $\sigma )$}\,, \label{M2_null}
\end{equation}%
\begin{equation}
    H_{1}:(\delta _{1},\,\delta _{2})=(\delta _{1}^{\ast },\,\delta _{2}^{\ast }) \hspace{0.25cm}\text{for some unknown $\delta _{1}^{\ast }\neq \delta_{2}^{\ast }$ }, \label{M2_alt}
\end{equation}%
where $\delta _{1}=(\mu _{1}$, $\sigma _{1})$ and\ $\delta _{2}=(\mu _{2}$, $\sigma _{2})$. Clearly, $H_{0}$ is a restricted version of $H_{1}$: for each $\theta _{0}\in \bar{\Omega}_{0}$, we can find $\theta _{1}$ such that 
\begin{equation}
    L_{T}^{0}(\theta _{0})=L_{T}(\theta _{1})\,,\quad \theta _{1}\in \Omega _{0},
\end{equation}%
where $\Omega _{0}$ is the subset of vectors $\theta _{1}\in \Omega $ such that $\theta _{1}$ satisfies $H_{0}$. Under $H_{0}$, the vector $\theta_{0}\in \bar{\Omega}_{0}$ is a nuisance parameter: the null distribution of any test statistic for $H_{0}$ depends on $\theta _{0}\in \bar{\Omega}_{0}$. In this problem, the null distribution of the test statistic, $LR_{T}$, is in fact completely determined by $\theta _{0}\in 
\bar{\Omega}_{0}$. As in \cite{garcia98} and the parametric bootstrap procedure describe in \cite{quzhuo2021likelihood} and \cite{kasshi2018}, it is assumed that the null hypothesis depends only on the mean, variance, and autoregressive coefficients. The likelihood ratio statistic for testing $H_{0}$ against $H_{1}$ can then written as 
\begin{equation}
    LR_{T}=2[\bar{L}_{T}(H_{1})-\bar{L}_{T}(H_{0})]
\end{equation}%
where 
\begin{equation}
    \bar{L}_{T}(H_{1})=\sup \{L_{T}(\theta _{1}):\theta _{1}\in \Omega \}\,,
\end{equation}%
\begin{equation}
    \bar{L}_{T}(H_{0})=\sup \{L_{T}^{0}(\theta _{0}):\theta _{0}\in \bar{\Omega}_{0}\}=\sup \{L_{T}(\theta _{1}):\theta _{1}\in \Omega _{0}\}\,.
\end{equation}
Since the model is parametric, we can generate a vector $N\;$i.i.d replications of $LR_{T}$ for any given value of $\theta _{0}\in \bar{\Omega}_{0}$:%
\begin{equation}
    LR(N,\,\theta _{0}):=[LR_{T}^{(1)}(\theta _{0}),\,\ldots\,,LR_{T}^{(N)}(\theta _{0})]^{\prime },\hspace{1cm}\theta _{0}\in \bar{\Omega}_{0}\,.
\end{equation}%
As discussed in \cite{rodrondufour_mcmstest}, the main assumptions required are that the random variables $LR_{T}^{(0)}, \,LR_{T}^{(1)}(\theta _{0}),\,\ldots \,,LR_{T}^{(N)}(\theta _{0})$ are exchangeable for some $\theta _{0}\in \bar{\Omega}_{0}$ each with distribution function $F[x\,|\,\theta _{0}]$ (i.e., they are $i.i.d$). From here, we can compute the Monte Carlo $p$-value which is given by  
\begin{equation}
    \hat{p}_{N}[x\,|\,\theta _{0}]=\frac{N+1-R_{LR}[LR_{T}^{(0)};\,N]}{N+1} \label{mmc_pval}
\end{equation}
where 
\begin{align}
    R_{LR}[LR_{T}^{(0)};\,N]& =\sum_{i=1}^{N}I[LR_{T}^{0}\geq LR_{T}^{i}(\theta _{0})]  \label{rank_fun}
\end{align}%
and $I(C):=1$ if condition $C$ holds, and $I(C)=0$ otherwise. As can be seen from (\ref{rank_fun}), $R_{LR}[LR_{T}^{(0)};$ \thinspace $N]$ simply computes the rank of the test statistic from the observed data within the generated series $LR(N,$\thinspace $\theta _{0})$. Then, as shown in \cite{rodrondufour_mcmstest}, a critical region for this test statistic with level $\alpha $ is then given by 
\begin{equation}
    \sup_{\theta _{0}\in \bar{\Omega}_{0}}\,\hat{p}_{N}[LR_{T}^{(0)}\,|\,\theta_{0}]\leq \alpha 
\end{equation}%
More precisely, if $(N+1)\alpha $ is an integer, then
\begin{equation}
    \mathbb{P}\left[ \sup \{\hat{p}_{N}[LR_{T}^{(0)}\,|\,\theta _{0}]:\theta_{0}\in \bar{\Omega}_{0}\}\leq \alpha \right] \leq \alpha 
\end{equation}%
under the null hypothesis and so it is a valid test with level $\alpha $ for $H_{0}$ and this result does not depend on the sample size $T$ and so it is also valid in finite samples.

This is the Maximized Monte Carlo likelihood ratio test and it requires searching for the maximum Monte Carlo p-value over the nuisance parameter space $\bar{\Omega}_{0}$. Since this space can be very large and grows as
the number of autoregressive components and the number of regimes increases, the authors propose a more efficient alternative which involves searching over a consistent set $C_T$ [as originally proposed in \cite{dufour2006}]. A consistent set can be defined using the consistent point estimate. For example, let $\hat{\theta}_{0}$ be the consistent point estimate of $\theta_{0}$. Then, we can define%
\begin{equation}
    C_{T}=\{\theta _{0}\in \bar{\Omega}_{0}:\left\Vert {}\right. \hat{\theta}_{0}-\theta _{0}\left. {}\right\Vert <c\}
\end{equation}%
where $c$ is a fixed positive constant that does not depend on $T$ and $\left\Vert \cdot \right\Vert $ is the Euclidean norm in $\mathbb{R}^{k}$. A consistent set of interest may also be $C_{T}^{\ast }=C_{T}^{CI}\cup C_{T}^{\epsilon }$ where 
\begin{align}
    C_{T}^{CI}& =\{\theta _{0}\in \bar{\Omega}_{0}:\left\Vert {}\right. \hat{\theta}_{0}-\theta _{0}\left. {}\right\Vert <2\times S.E.(\hat{\theta}_{0})\}\\
    C_{T}^{\epsilon }& =\{\theta _{0}\in \bar{\Omega}_{0}:\left\Vert {}\right. \hat{\theta}_{0}-\theta_{0}\left. {}\right\Vert <\epsilon \}
\end{align}%
Hence, $C_{T}^{CI}$ is defined by a confidence interval based on consistent point estimates, while $C_{T}^{\epsilon }$ is determined using a fixed constant $\epsilon$ that is independent of $T$. The union of these two sets allows for values that may lie outside the confidence interval for some parameters and within it for others, depending on the choice of $\epsilon$. The \pkg{MSTest} \proglang{R} package enables users to define the consistent set $C_T$ by specifying only a fixed positive constant $\epsilon$, only the confidence interval, or the union of both. 

As discussed in \cite{dufour2006} and \cite{rodrondufour_mcmstest}, the solution to this optimization problem may not be unique, meaning that the maximum $p$-value may correspond to multiple parameter vectors. For this reason, derivative-free numerical optimization methods are recommended to locate the maximum Monte Carlo $p$-value within the nuisance parameter space. The package \pkg{MSTest} allows users to use the Generalized Simulated Annealing algorithm, Genetic Algorithm, and Particle Swarm algorithm [see  \cite{xiaetal2013}, \cite{zametal2013}, \cite{scrucca2013}, \cite{dufour2006}, and \cite{dufour2019finite}].

Finally, as described in \cite{rodrondufour_mcmstest}, we can define $C_{T}$ as the singleton set $C_{T}={\hat{\theta}_{0}}$, resulting in the Local Monte Carlo Likelihood Ratio Test (LMC-LRT). Here, the consistent set includes only the consistent point estimate $\hat{\theta}_{0}$, so the Monte Carlo $p$-value depends solely on $\hat{\theta}_{0}$. This LMC version of the test can be interpreted as the finite-sample analogue of the parametric bootstrap. In this context, asymptotic validity pertains to $\hat{\theta}_{0}$ converging to the true parameter $\theta _{0}$ as the sample size increases, rather than to the asymptotic validity of critical values emphasized in studies like \cite{hansen92}, \cite{garcia98}, \cite{chowhite07}, \cite{quzhuo2021likelihood}, and \cite{kasshi2018}. Specifically, akin to the parametric bootstrap, the LMC procedure is valid only as $T\rightarrow \infty$. However, unlike the parametric bootstrap, a large number of simulations (\emph{i.e.}, $N\rightarrow \infty$) is not necessary, as the procedure does not approximate asymptotic critical values or assume asymptotic convergence of the test statistic distribution but instead uses critical values derived from the sample distribution. This design allows for computational efficiency since it eliminates the need for extensive simulations to obtain asymptotically valid critical values. Further, as discussed in \cite{rodrondufour_mcmstest}, The procedure is valid even in cases where an asymptotic distribution does not exist. This is another feature which makes both the MMC-LRT and the LMC-LRT procedures more general than the parametric bootstrap. Still, it is worth noting that the parametric bootstrap procedure discussed in \cite{quzhuo2021likelihood} and \cite{kasshi2018} can be implemented by constraining the parameter space of the transition probabilities based on the assumptions in those studies. Using a larger number of simulations in these contexts can approximate asymptotic critical values where prior research has demonstrated the parametric bootstrap’s validity. Nevertheless, as outlined in \cite{rodrondufour_mcmstest}, the LMC-LRT and MMC-LRT procedures remain the most general likelihood ratio test procedures currently available.
% ----------------------------------------- %
% ----- Moment-based test
% ----------------------------------------- %
\subsection{Moment-based tests}
\cite{dufourluger17} propose a different way to test Markov switching models that also avoids the statistical issues described above for likelihood ratio type tests. Additionally, their test is less costly computationally in comparison to all test mentioned above, the parameter stability test discussed next, and allows the econometrician to perfectly control the level of the test through the use of the Monte Carlo test methods described in \cite{dufour2006} and used in \cite{rodrondufour_mcmstest}. However, their proposed method can only deal with the case where $M_0=m=1$. The moment-based test of \cite{dufourluger17} involves computing  moments of the least-square residuals of autoregressive models under the null hypothesis. More Specifically, they focus on the mean, variance, skewness and excess kurtosis of the least-square residuals. These moments are calculate as
\begin{equation}
    M(\hat{\epsilon}) = \frac{|m_2 - m_1|}{\sqrt{s^2_1+s^2_2}}
\end{equation}
\noindent
where, \(m_1 = \frac{\Sigma^T_{t=1} \hat{\epsilon_t} \mathbbm{1}[\hat{\epsilon_t} < 0]}{\Sigma^T_{t=1}\mathbbm{1}[\hat{\epsilon_t} < 0]}\), \(m_2 = \frac{\Sigma^T_{t=1} \hat{\epsilon_t} \mathbbm{1}[\hat{\epsilon_t} > 0]}{\Sigma^T_{t=1}\mathbbm{1}[\hat{\epsilon_t} > 0]}\), \(s^2_1 = \frac{\Sigma^T_{t=1} (\hat{\epsilon_t}-m_1)^2 \mathbbm{1}[\hat{\epsilon_t} < 0]}{\Sigma^T_{t=1}\mathbbm{1}[\hat{\epsilon_t} < 0]}\) and \(s^2_2 = \frac{\Sigma^T_{t=1} (\hat{\epsilon_t}-m_2)^2 \mathbbm{1}[\hat{\epsilon_t} > 0]}{\Sigma^T_{t=1}\mathbbm{1}[\hat{\epsilon_t} > 0]}\)
\begin{equation}
    V(\hat{\epsilon}) = \frac{\vartheta_2(\hat{\epsilon})}{\vartheta_1(\hat{\epsilon})}
\end{equation}
\noindent
where, \(\vartheta_1 = \frac{\Sigma^T_{t=1} \hat{\epsilon_t}^2 \mathbbm{1}[\hat{\epsilon_t}^2 < \hat{\sigma}^2]}{\Sigma^T_{t=1} \mathbbm{1}[\hat{\epsilon_t}^2 < \hat{\sigma}^2]} \), \(\vartheta_2 = \frac{\Sigma^T_{t=1} \hat{\epsilon_t}^2 \mathbbm{1}[\hat{\epsilon_t}^2 > \hat{\sigma}^2]}{\Sigma^T_{t=1} \mathbbm{1}[\hat{\epsilon_t}^2 > \hat{\sigma}^2]} \) and \(\hat{\sigma}^2=T^{-1}\Sigma^T_{t=1} \hat{\epsilon_t}^2\)
\begin{equation}
    S(\hat{\epsilon})= | \frac{\Sigma^T_{t=1} \hat{\epsilon_t}^3}{T(\hat{\sigma}^2)^{3/2}} |
\end{equation}
\noindent
and
\begin{equation}
    K(\hat{\epsilon})= | \frac{\Sigma^T_{t=1} \hat{\epsilon_t}^4}{T(\hat{\sigma}^2)^{2}} -3|
\end{equation}
\noindent
The testing procedure involves calculating the test statistic for each moment, obtaining the individual p-values and using two different methods of combining independent test statistics. The first method is based on the min of the p-values and was suggested by \cite{tippett1931} and \cite{wilkinson1951}. Here, the test statistic becomes,
\begin{equation}
    F_{min}(\hat{\epsilon}) = 1 - min\{\hat{G_M}[M(\hat{\epsilon})],\hat{G_V}[V(\hat{\epsilon})],\hat{G_S}[S(\hat{\epsilon})],\hat{G_K}[K(\hat{\epsilon})]\}
\end{equation}
\noindent
where for example, \(\hat{G_M}[M(\hat{\epsilon})] = 1 - \hat{F_M}[M(\hat{\epsilon})]\) is the Monte Carlo p-value of \(M(\hat{\epsilon})\). The second method of combining the test statistics involves taking the product of them. This method of combining test statistics was suggested by \cite{fisher1932} and \cite{pearson1933}. In this case the the test statistic becomes,
\begin{equation}
    F_{\times}(\hat{\epsilon}) = 1 - \{\hat{G_M}[M(\hat{\epsilon})] \times \hat{G_V}[V(\hat{\epsilon})] \times \hat{G_S}[S(\hat{\epsilon})] \times \hat{G_K}[K(\hat{\epsilon})]\}
\end{equation}
\noindent
Interested readers should see \cite{dufour2004} and \cite{dufour2014} which provide further discussion of these methods of combining test statistics. Finally, the Monte Carlo p-value of the combined statistics is given by
\begin{equation}
    G_{F_{min}}[F_{min}(\hat{\epsilon});N] = \frac{N+1-R_{F_{min}}[F_{min}(\hat{\epsilon});N]}{N}
\end{equation}
\noindent
and
\begin{equation}
    G_{F_{\times}}[F_{\times}(\hat{\epsilon});N] = \frac{N+1-R_{F_{\times}}[F_{\times}(\hat{\epsilon});N]}{N}
\end{equation}
\noindent
where \(R_{F_{min}}\) and \(R_{F_{\times}}\) are the ranks of \(F_{min}(\hat{\epsilon})\) and \(F_{\times}(\hat{\epsilon})\) in  \(F_{min}(\hat{\eta}_1)\), ..., \(F_{min}(\hat{\eta}_{N-1})\) and \(F_{\times}(\hat{\eta}_1),\) ...,\( F_{\times}(\hat{\eta}_{N-1})\)    respectively, when ordered. Also, \(\hat{\eta} = \eta - \bar{\eta}\) and \(\eta \sim N(0,I_T)\).

The computational efficiency of this test makes it easily extendable to the use of Maximized Monte Carlo when nuisance parameters are present. Furthermore, this test is not subject to the same level of statistical difficulties such as unidentified parameters under the null as in \cite{hansen92}, \cite{garcia98} and \cite{chp14}. This is because, transition probability parameters, the mean and the variance do not need to be treated as nuisance parameters. Parameters of explanatory variables are the only ones which may be unidentified under the null and so only these are treated as nuisance parameters. \cite{garcia98} also reduced the nuisance parameter space by treating only the transition probabilities $p_{11}$ and $q_{22}$ as nuisance parameters, however, this is an even further reduction of the nuisance parameter space and makes the treatment of autoregressive models with more lags more tractable. Although the moment-based test can only be used to compare linear models against Markov switching models with two regimes, it is the least computationally intensive procedure available and takes only seconds to compute, even when considering the MMC version of the test, and so it is included in the \proglang{R} package \pkg{MSTest}. 

% ----------------------------------------- %
% ----- Parameter stability test
% ----------------------------------------- %
\subsection{Optimal test for regime switching}
\cite{chp14} proposes a test that can be described as an optimal test for the consistency of parameters in random coefficient and Markov switching models. This test can be understood as an extension of \cite{white1982}'s information matrix test. As suggested by the authors, it shares several advantages, such as the need to estimate the model only under the null hypothesis, which, as we saw, is also the case for the moment-based test of \cite{dufourluger17}. This feature of needing to estimate only the restricted model is particularly advantageous. In contrast, the likelihood ratio tests proposed by \cite{hansen92}, \cite{garcia98}, \cite{chowhite07}, \cite{quzhuo2021likelihood}, \cite{kasshi2018}, and \cite{rodrondufour_mcmstest} all require estimating the model under both the null and alternative hypotheses. The presence of non-linearity in estimating Markov switching models introduces multiple local optima, necessitating numerical procedures and making likelihood ratio test procedures relatively more computationally intensive. Moreover, the authors invoke the Neyman-Pearson lemma to prove the optimality of their test, demonstrating that it is asymptotically locally equivalent to the likelihood ratio test. However, simulation results presented in \cite{rodrondufour_mcmstest} and discussion in \cite{quzhuo2021likelihood} suggest their LRT approach have better power in certain settings, such as when only the mean is subject to change and persistence is high. Further, this method also involves a bootstrap procedure while searching over the nuisance parameter space, which can make obtaining asymptotic critical values more computationally intensive than the moment-based approach of \cite{dufourluger17}, for example.

The authors formulate the hypothesis in the following way:
\begin{equation}
    \begin{aligned}
        H_0: \theta_t &= \theta_0 \\
        H_1: \theta_t &= \theta_0 + \eta_t
    \end{aligned}
\end{equation}
where the switching variable \(\eta_t\) is unobservable, stationary and may depend on nuisance parameters \(\beta\). Their test makes use of the second derivatives of the log-likelihood and the outer products of the scores as in the information matrix test with the addition of an extra term, which captures the serial dependence of the time-varying coefficients. This means that the form of the test depends on the latent process \(\eta_t\) only through its second-order properties. Additionally, the distribution of \( \eta_t \) is assumed to exist even under the null, but does not play a role with regards to the distribution of the data \((y_T, y_{T-1},y_{T-2},...y_1)\) under the null. That is, under the null, they are mutually exclusive. 

The authors first propose a Sup-type test as in \cite{davies87} to combat the presence of nuisance parameters. They set \(\eta_t = chS_t \), where $c$ is a scalar specifying the amplitude of the change, $h$ a vector specifying the direction of the alternative and \(S_t\) is a Markov-chain, which follows an autoregressive process such as \(S_t = \rho S_{t-1} + e_t\), where \(e_t\) is i.i.d. U[-1,1] and \(-1 <\rho < 1\) so that \(S_t\) is bounded by support \(( -1/(1-|\rho|), 1/(1-|\rho|)\) and has zero mean.  Letting \(\beta =(c^2, h',\rho ') \) be the vector of nuisance parameters, we can write
\begin{equation}
    \mu_{2,t}(\beta,\theta) = \frac{1}{2}c^2 h'[(\frac{\partial l_t}{\partial\theta\partial\theta'}+(\frac{\partial l_t}{\partial\theta}) (\frac{\partial l_t}{\partial\theta})') + 2 \sum\limits_{s<t}\rho^{(t-s)}(\frac{\partial l_t}{\partial\theta}) (\frac{\partial l_t}{\partial\theta})']h
\end{equation}
\noindent
which allows us to get the expression
\begin{equation}
    \text{supTS} = \sup_{\{h, p:||h||=1, \underline{\rho} < \rho < \bar{\rho}\}} = \frac{1}{2} (max(0, \frac{\Gamma^*_T}{\sqrt{\hat{\epsilon^{*'}}\hat{\epsilon^*}}}))^2
\end{equation}
as in \cite{chp14}, where \(\mu^*_{2,t}(\beta,\theta) = \mu_{2,t}(\beta,\theta)/c^2\), \(\Gamma^*_T = \Gamma^*_T(\beta,\theta) = \sum\limits_{t} \mu^*_{2,t}(\beta,\theta)/\sqrt{T}\) and \(\hat{\epsilon^*}\) are the residuals from regressing \(\mu^*_{2,t}(\beta,\theta)\) on \( l_t^{(1)}(\hat{\theta})\) so that \(\Gamma^*_T \) and \(\hat{\epsilon^*}\) are not dependent on \(c^2\). As previously mentioned, this methodology involves bootstrapping over the distributions of the nuisance parameters. As a result, one must choose a prior distribution for the nuisance parameters. The most commonly used distribution in this case is the uniform distribution, and this is what is implemented in the \pkg{MSTest} package. However, since the parameter \(c^2\) is not necessarily bounded from above, a uniform distribution may not always be appropriate. As a result, \cite{chp14} also suggest using an Exponential-type test as in \cite{andrewsploberger94}. They propose the following statistic:
\begin{equation}
    expTS = \int\limits_{\{\underline{\rho} \leq \rho \leq \bar\rho, ||h||<1 \}} \Psi(h,\rho) d\rho dh
\end{equation}
where 
\begin{equation}
    \Psi(h,\rho) = 
    \begin{cases}
     	\sqrt{2 \pi} exp[\frac{1}{2}(\frac{\Gamma^*_T}{\sqrt{\hat{\epsilon^{*'}}\hat{\epsilon^*}}} - 1 )^2]	\Phi (\frac{\Gamma^*_T}{\sqrt{\hat{\epsilon^{*'}}\hat{\epsilon^*}}} - 1 ) & \text{if \(\hat{\epsilon^{*'}}\hat{\epsilon^*} \neq 0. \)} \\
       	1 	& \text{otherwise.}
   	\end{cases}
\end{equation}

These tests proposed by \cite{chp14} have been used in the empirical applications of \cite{hamilton05}, \cite{warnevredin06}, \cite{kahnrich07}, \cite{morleypiger12}, and \cite{DMPF11}, in testing MS-GARCH models by \cite{hushin08}, and in \cite{dufourluger17}, \cite{quzhuo2021likelihood}, and \cite{rodrondufour_mcmstest} as a benchmark to compare their test procedures. Due to their wide use and optimality results, the \pkg{MSTest} package also includes these test procedures.

%% -- software ----------------------------------------------------------------

%% - In principle "as usual" again.
%% - When using equations (e.g., {equation}, {eqnarray}, {align}, etc.
%%   avoid empty lines before and after the equation (which would signal a new
%%   paragraph.
%% - When describing longer chunks of code that are _not_ meant for execution
%%   (e.g., a function synopsis or list of arguments), the environment {Code}
%%   is recommended. Alternatively, a plain {verbatim} can also be used.
%%   (For executed code see the next section.)

\section{The R package MSTest} \label{sec:software}
The MSTest package is designed for conducting hypothesis testing for the number of regimes in Markov switching models within the \proglang{R} programming environment. Since many of the test procedures included here require estimating the restricted and unrestricted models and simulating the restricted model, it also makes available simulation and estimation of Markov switching models. However, it should be noted that these features are simply a by-product and not the focus of this package. Specifically, they were designed with the purpose of being compatible with the hypothesis testing functions. 

Several other platforms also offer functionalities for Markov switching models. for example, in \proglang{MATLAB}, the Econometrics Toolbox provides functions for estimating state-space models, including Kalman filters, though it may lack hypothesis testing for these models. In \proglang{Python}, the \pkg{statsmodels} library also offers support for estimating state-space models, including Markov switching models, but again does not include dedicated functions for hypothesis testing of Markov switching model. Similarly, in \proglang{STATA}, while commands like \code{mswitch} exist for estimating Markov switching models, they do not cover the extensive range of testing procedures that \pkg{MSTest} provides.

For these reasons, \pkg{MSTest} stands out in its implementation of novel testing procedures that are robust to the violation of regularity condition. Additionally, the integration of Rcpp enhances computational efficiency, allowing users to handle nuisance parameters effectively. Hence, while there are alternative tools available across platforms for simulating and estimating Markov switching models, \pkg{MSTest} offers a unique combination of flexibility, computational efficiency, and availability of hypothesis testing procedures that sets it apart.

% ----------------------------------------- %
% ----- Data sets
% ----------------------------------------- %
\subsection{Data sets}
The R package \pkg{MSTest} includes three samples of U.S. real GNP and one sample of U.S. real GDP, all of which can easily be accessed once the package is loaded. Specifically, it provides the original sample used in \cite{hamilton89}, the sample ending in 2010 first considered in \cite{chp14}, and a more complete sample ranging from the second quarter of 1947 to the second quarter of 2024. The U.S. real GDP data also covers this extended period. These samples have been used by \cite{hansen92}, \cite{chp14}, \cite{dufourluger17}, and \cite{rodrondufour_mcmstest}, among others, to test for the number of regimes in a Markov switching model and to showcase the performance of their proposed test procedures. Table \ref{tab:datasets} provides the label used to identify each sample in the \proglang{R} package \pkg{MSTest} and describes the span of each sample.
\begin{table}[!tbh]
\centering
\begin{tabular}{p{0.2\textwidth}  p{0.7\textwidth}}
    \hline
    Label & Descriptions \\
    \hline
    \code{hamilton84GNP} & Sample originally considered in \cite{hamilton89}.  This series ranges from 1951Q2 to 1984Q4.\\
    \code{chp10GNP} & This sample is the second sample of the U.S. real GNP used in \cite{chp14} and \cite{dufourluger17}. This series ranges from 1951Q2 to 2010Q4.\\
    \code{USGNP} & This sample is used in \cite{rodrondufour_mcmstest}. The series ranges from 1947Q2 to 2024Q2.\\
    \code{USRGNP} & This sample is used in \cite{rodrondufour_mcmstest}. The series ranges from 1947Q2 to 2024Q2.\\
    \hline
    \end{tabular}
\caption{\label{tab:datasets} U.S. real GNP samples in \pkg{MSTest}}
\end{table} 

These data sets can be loaded using the following commands in \proglang{R} once \pkg{MSTest} has been loaded: 
\begin{CodeChunk}
\begin{CodeInput}
R> data("hamilton84GNP", package = "MSTest")
R> data("chp10GNP", package = "MSTest")
R> data("USGNP", package = "MSTest")
R> data("USRGDP", package = "MSTest")
\end{CodeInput}
\end{CodeChunk}
All three data sets include three columns: (1) \code{Date}, (2) \code{GNP} or \code{RGDP}, and (3) \code{GNP\_gr} or \code{RGDP\_gr}. The first column is of \code{Date} type, defined using the \code{as.Date()} function. The second column contains the levels of U.S. real GNP or GDP, and the third column contains the growth rate of U.S. real GNP or GDP.
% ----------------------------------------- %
% ----- Simulation
% ----------------------------------------- %
\subsection{Simulation}
This section describes a set of functions available in \pkg{MSTest} that allow users to simulate different types of processes. These functions are utilized by the hypothesis testing procedures, specifically those that involve using simulation to obtain the (sample or asymptotic) null distribution of the test statistic, such as \code{LMCLRTest}, \code{MMCLRTest}, \code{DLMCTest}, \code{DLMMCTest}, and \code{CHPTest}. Users interested in developing new estimation or testing procedures for Markov switching models may find these functions useful for testing the performance of their procedures and comparing them to those available in \pkg{MSTest}.

Table \ref{tab:simulation} provides the labels used to identify each simulation function in the \pkg{MSTest} package and describes the processes they can simulate. Each function requires a \code{List} object as input, containing values for the data-generating process. Below, we present examples for some of the included simulation functions, which should give users an idea of the elements the input \code{List} must include. As with all functions in \pkg{MSTest}, a more exhaustive description of each function's inputs is provided in the \pkg{MSTest} documentation available through CRAN.

\begin{table}[!tbh]
\centering
\begin{tabular}{p{0.2\textwidth}  p{0.7\textwidth} }
    \hline
    Function & Description \\
    \hline
    \code{simuNorm} & Simulates a normally distributed process. Can also include $X$ exogenous regressors.\\
    \code{simuAR}  & Simulate an autoregressive process with $p$ lags (AR($p$)).\\
    \code{simuARX}  & Simulate an autoregressive process with $p$ lags (AR($p$)) and $X$ as exogenous regressors.\\
    \code{simuVAR} & Simulate a vector autoregressive process with $p$ lags (VAR($p$)). \\
    \code{simuVARX} & Simulate a vector autoregressive process with $p$ lags (VAR($p$)) and $X$ as exogenous regressors.\\
    \code{simuMSAR}  & Simulate a Markov switching autoregressive process with $p$ lags (MSAR($p$)).\\
    \code{simuMSARX}  & Simulate a Markov switching autoregressive process with $p$ lags (MSAR($p$)) and $X$ as exogenous regressors.\\
    \code{simuMSVAR}  & Simulate a Markov switching vector autoregressive process with $p$ lags (MSVAR($p$)). \\
    \code{simuMSVARX}  & Simulate a Markov switching vector autoregressive process with $p$ lags (MSVAR($p$)) and $X$ as exogenous regressors.\\
    \code{simuHMM}  & Simulate a hidden Markov process (HMM). Can also include $X$ exogenous regressors.\\
    \hline
    \end{tabular}
\caption{\label{tab:simulation} Simulation functions available in the \proglang{R} package \pkg{MSTest}}
\end{table}

For example, as described in Table \ref{tab:simulation}, simulating a normally distributed process can be done using the \code{simuNorm} function. In this case, the function requires the user to input a \code{List} containing: the sample size (\code{n}), the number of series to be simulated (\code{q})—where \code{q=1} indicates a univariate process and \code{q>1} indicates a multivariate process—a ($q \times 1$) vector of means for each series, and a ($q \times q$) covariance matrix. Users may also specify the number of additional observations to simulate and discard at the beginning using the \code{burnin} parameter, which for \code{simuNorm} defaults to $0$. For autoregressive type processes however, the default \code{burnin} value is higher to avoid any dependence on the initialization used in simulating the process. Users also have the option to provide a ($(n+\text{burnin}) \times q$) matrix of errors if they prefer not to use normally distributed errors. This is done by defining the element \code{eps} in the input \code{list} with that matrix. As an example, we can simulate a multivariate normal process, an autoregressive process, a vector autoregressive process, a Markov switching autoregressive process, a hidden Markov process, and a Markov switching vector autoregressive process using the following code:

\begin{CodeChunk}
\begin{CodeInput}
R> # Define DGP of multivariate normal process
R> mdl_norm <- list(n     = 500, 
+                   q     = 2,
+                   mu    = c(5, -2),
+                   sigma = rbind(c(5.0, 1.5),
+                                 c(1.5, 1.0)))
R> # Simulate process 
R> simu_norm <- simuNorm(mdl_norm)
R> 
R> # Define DGP of AR(2) process
R> mdl_ar <- list(n     = 500, 
+                 mu    = 5,
+                 sigma = 1,
+                 phi   = c(0.75))
R> # Simulate process 
R> simu_ar <- simuAR(mdl_ar)
R> 
R> # Define DGP of VAR(2) process
R> mdl_var <- list(n     = 500, 
+                  p     = 1,
+                  q     = 2,
+                  mu    = c(5, -2),
+                  sigma = rbind(c(5.0, 1.5),
+                                c(1.5, 1.0)),
+                  phi   = rbind(c(0.50, 0.30),
+                                c(0.20, 0.70)))
R> # Simulate process 
R> simu_var <- simuVAR(mdl_var)
R> 
R> 
R> # Define DGP of HMM
R> mdl_hmm <- list(n     = 500, 
+                  q     = 2,
+                  mu    = rbind(c(5, -2),
+                                c(10, 2)),
+                  sigma = list(rbind(c(5.0, 1.5),
+                                     c(1.5, 1.0)),
+                               rbind(c(7.0, 3.0),
+                                     c(3.0, 2.0))),
+                  k     = 2,
+                  P     = rbind(c(0.95, 0.10),
+                                c(0.05, 0.90)))
R> # Simulate process 
R> simu_hmm <- simuHMM(mdl_hmm)
R> 
R> # Define DGP of MS AR process
R> mdl_ms <- list(n     = 500, 
+                 mu    = c(5,10),
+                 sigma = c(1,2),
+                 phi   = c(0.75),
+                 k     = 2,
+                 P     = rbind(c(0.95, 0.10),
+                               c(0.05, 0.90)))
R> # Simulate process 
R> simu_msar <- simuMSAR(mdl_ms)
R> 
R> # Define DGP of MS VAR process
R> mdl_msvar <- list(n     = 500, 
+                    p     = 1,
+                    q     = 2,
+                    mu    = rbind(c(5, -2),
+                                  c(10, 2)),
+                    sigma = list(rbind(c(5.0, 1.5),
+                                       c(1.5, 1.0)),
+                                 rbind(c(7.0, 3.0),
+                                       c(3.0, 2.0))),
+                    phi   = rbind(c(0.50, 0.30),
+                                  c(0.20, 0.70)),
+                    k     = 2,
+                    P     = rbind(c(0.95, 0.10),
+                                  c(0.05, 0.90)))
R> # Simulate process
R> simu_msvar <- simuMSVAR(mdl_msvar)
\end{CodeInput}
\end{CodeChunk}
 
\begin{figure}[!tbh]
\centering
\includegraphics{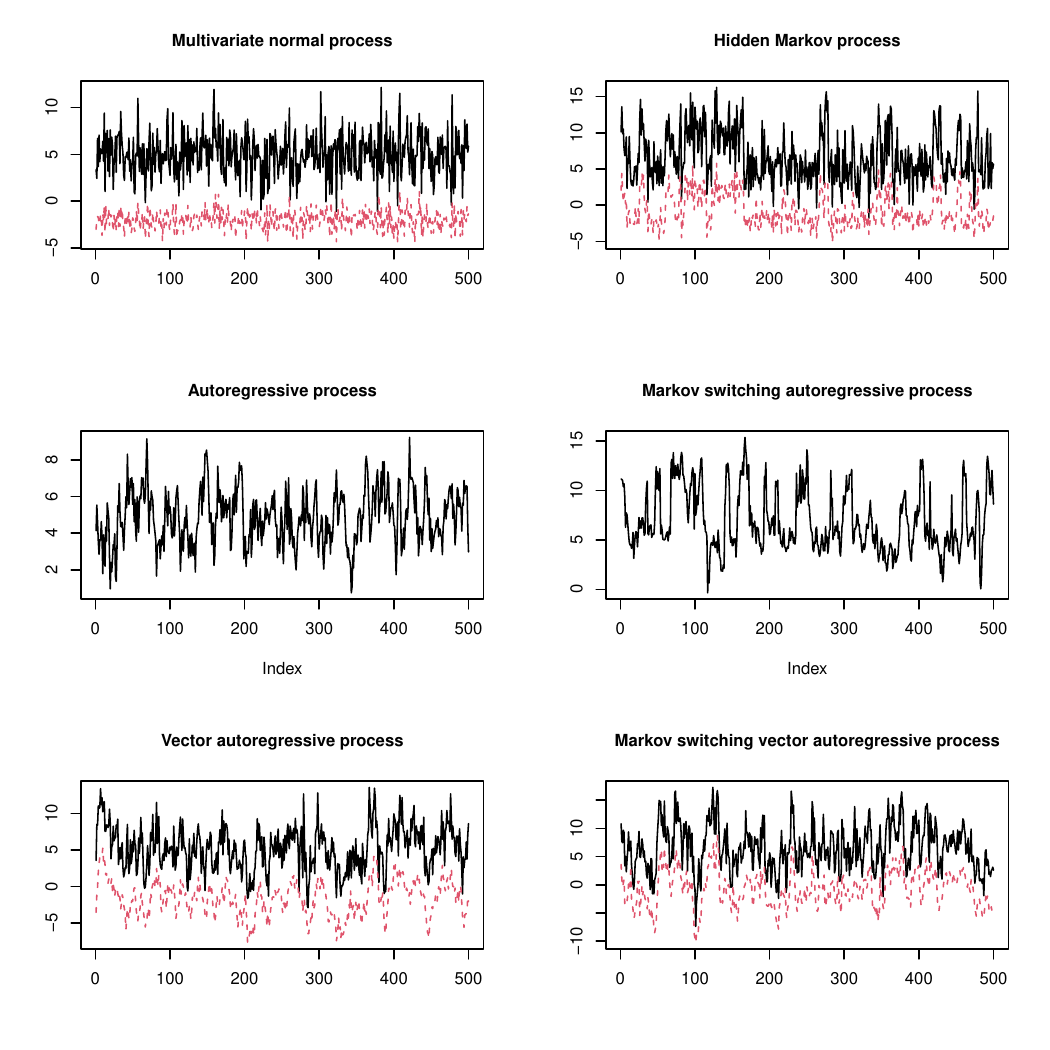}
\caption{\label{fig:sims} Simulated processes. Linear process on left and Markov switching process on right.}
\end{figure}

These simulated processes are shown in Figure \ref{fig:sims}. From here, we can already observe the regime switching nature of some of the processes. It is especially more obvious when comparing the two middle charts which plot an autoregressive process (left) and Markov switching autoregressive process (right).
% ----------------------------------------- %
% ----- Model estimation
% ----------------------------------------- % 
\subsection{Model estimation}
Here, we briefly describe the set of functions available in \pkg{MSTest} that allow users to estimate different types of Markov switching models. In total, there are ten distinct types of models that can be estimated. These models are listed in Table \ref{tab:models}, along with their labels in \pkg{MSTest} and the corresponding equations for each model. Although \code{Nmdl} and \code{HMmdl} use multivariate notation, if a ($T \times 1$) vector is provided as input, the function will detect that this is a univariate setting. Also, as may be apparent from their equations, the labels that include an `X' in their names are those that allow for the inclusion of exogenous regressors. Although the \code{Nmdl} and \code{HMmdl} functions do not follow this nomenclature, including exogenous regressors within these functions is always possible.

\begin{table}[!tbh]
\centering
\begin{tabular}{p{0.2\textwidth}  p{0.2\textwidth} p{0.5\textwidth}}
    \hline
    Model & Label & Equation \\
    \hline
    N$(\pmb{\mu} +  \pmb{x}_{t} \pmb{\beta}, \pmb{\Sigma})$ & \code{Nmdl} & $\pmb{y}_t = \pmb{\mu} +  \pmb{x}_{t} \pmb{\beta} + \pmb{\Sigma}^{1/2} \pmb{\epsilon}_t$ \\
    AR($p$) & \code{ARmdl} & $y_t = \mu + \sum^p_{k=1} \phi_k (y_{t-k} - \mu) + \sigma \epsilon_t$ \\
    ARX($p$) & \code{ARXmdl} & $y_t = \mu + \sum^p_{k=1} \phi_k (y_{t-k} - \mu) +  \pmb{x}_{t} \pmb{\beta} + \sigma \epsilon_t$ \\
    VAR($p$) & \code{VARmdl} & $\pmb{y}_t  = \pmb{\mu} + \sum^p_{k=1}(\pmb{y}_{t-k} - \pmb{\mu})\pmb{\Phi}_k + \pmb{\Sigma}^{1/2}\pmb{\epsilon}_t$\\
    VARX($p$) & \code{VARXmdl} & $\pmb{y}_t  = \pmb{\mu} + \sum^p_{k=1}(\pmb{y}_{t-k} - \pmb{\mu})\pmb{\Phi}_k + \pmb{x}_t \pmb{\beta} + \pmb{\Sigma}^{1/2}\pmb{\epsilon}_t$\\
    MS-AR($p$) & \code{MSARmdl} & $y_t = \mu_{s_t} + \sum^p_{k=1} \phi_k (y_{t-k} - \mu_{s_t}) + \sigma_{s_t} \epsilon_t$ \\
    MS-ARX($p$) & \code{MSARXmdl} & $y_t = \mu_{s_t} + \sum^p_{k=1} \phi_k (y_{t-k} - \mu_{s_t})  +  \pmb{x}_{t} \pmb{\beta} + \sigma_{s_t} \epsilon_t$ \\
    MS-VAR($p$) & \code{MSVARmdl} & $\pmb{y}_t  = \pmb{\mu}_{s_t} + \sum^p_{k=1}\pmb{\Phi}_k(\pmb{y}_{t-k} - \pmb{\mu}_{s_{t-k}}) + \pmb{\Sigma}^{1/2}_{s_t}\pmb{\epsilon}_t$\\
    MS-VARX($p$) & \code{MSVARXmdl} & $\pmb{y}_t  = \pmb{\mu}_{s_t} + \sum^p_{k=1}\pmb{\Phi}_k(\pmb{y}_{t-k} - \pmb{\mu}_{s_{t-k}}) +  \pmb{x}_{t} \pmb{\beta} + \pmb{\Sigma}^{1/2}_{s_t}\pmb{\epsilon}_t$\\
    HMM & \code{HMmdl} & $\pmb{y}_t  = \pmb{\mu}_{s_t} +  \pmb{x}_{t} \pmb{\beta} + \pmb{\Sigma}^{1/2}_{s_t}\pmb{\epsilon}_t$\\
    \hline
    \end{tabular}
\caption{\label{tab:models} Models and their specifications available in the \proglang{R} package \pkg{MSTest}}
\end{table}

Above, we used the simulation functions to simulate a multivariate Hidden Markov process with $q=2$ series, a Markov switching autoregressive process with $p=1$ lag, and a Markov switching vector autoregressive process with $p=1$ lag and $q=2$ series. The output of these functions is a list that includes the simulated process, the true state variable $S_{t}$, and other elements of the data-generating process (DGP) that were provided as input. Below, we provide an example where the simulated processes are used as input for estimating the respective model. This is demonstrated for the Hidden Markov model and the two Markov switching models.

\begin{CodeChunk}
\begin{CodeInput}
R> # Estimate Hidden Markov model 
R> # Set options for model estimation
R> control <- list(msmu   = TRUE, 
+                  msvar  = TRUE,
+                  method = "EM",
+                  use_diff_init = 30)
R> # Estimate model
R> mdl_est_hmm <- HMmdl(simu_hmm[["y"]], k = 2, control = control)
R> summary(mdl_est_hmm)

Hidden Markov Model
              coef     s.e.
mu_1,1    4.838100 0.124640
mu_2,1   -2.108900 0.058911
mu_1,2   10.188000 0.211110
mu_2,2    2.020600 0.119600
sig_11,1  4.890500 0.387420
sig_12,1  1.419100 0.153010
sig_22,1  1.021900 0.087453
sig_11,2  6.781900 0.782320
sig_12,2  2.926300 0.394830
sig_22,2  2.030200 0.243440
p_11      0.946270 0.051969
p_12      0.053734 0.013297
p_21      0.109500 0.026341
p_22      0.890500 0.073539

log-likelihood =  -1854.11
AIC =  3736.22
BIC =  3795.225

Residuals:
       Min       1Q     Median      3Q    Max
Y1 -6.6190 -1.62470 -0.0019167 1.42860 6.6616
Y2 -3.0745 -0.76757 -0.0090576 0.71458 3.7755
R> 
R> # Estimate Markov switching autoregressive model
R> # Set options for model estimation
R> control <- list(msmu   = TRUE, 
+                  msvar  = TRUE, 
+                  method = "EM",
+                  use_diff_init = 30)
R> # Estimate model
R> mdl_est_msar <- MSARmdl(simu_msar[["y"]], p = 1, k = 2, control = control)
R> summary(mdl_est_msar)

Markov Switching Autoregressive Model
           coef     s.e.
mu_1  10.754000 0.298580
mu_2   5.246300 0.226300
phi_1  0.788720 0.026999
sig_1  1.935100 0.231210
sig_2  0.937860 0.071588
p_11   0.880750 0.079630
p_12   0.119250 0.029808
p_21   0.050742 0.012311
p_22   0.949260 0.052574

log-likelihood =  -863.7257
AIC =  1745.452
BIC =  1783.365

Residuals:
       Min       1Q Median      3Q    Max
Y1 -3.8936 -0.70633 0.0393 0.77594 3.4368
R> 
R> # Estimate Markov switching vector autoregressive model 
R> # Set options for model estimation
R> control <- list(msmu   = TRUE, 
+                  msvar  = TRUE,
+                  method = "EM",
+                  use_diff_init = 30)
R> # Estimate model
R> mdl_est_msvar <- MSVARmdl(simu_msvar[["y"]], p = 1, k = 2, control = control)
R> summary(mdl_est_msvar)

Markov Switching Vector Autoregressive Model
              coef      s.e.
mu_1,1    9.003000 0.5299100
mu_2,1    0.949350 0.4783700
mu_1,2    4.829800 0.5036300
mu_2,2   -1.973600 0.4985200
phi_1,11  0.469290 0.0596180
phi_1,12  0.332650 0.0692590
phi_1,21  0.218650 0.0284450
phi_1,22  0.683660 0.0314900
sig_11,1  7.255500 0.8579700
sig_12,1  2.778400 0.4205600
sig_22,1  2.016000 0.1638000
sig_11,2  5.039900 0.4544000
sig_12,2  1.519000 0.1826100
sig_22,2  1.081900 0.1004800
p_11      0.856940 0.0050732
p_12      0.143060 0.0293330
p_21      0.091072 0.0190220
p_22      0.908930 0.0342020

log-likelihood =  -1917.78
AIC =  3871.561
BIC =  3947.388

Residuals:
       Min       1Q    Median      3Q    Max
Y1 -7.0841 -1.51650 0.0349380 1.59320 6.9019
Y2 -4.4121 -0.70895 0.0021277 0.71763 3.6255
\end{CodeInput}
\end{CodeChunk}

In each case, we set \code{method="EM"} in the \code{control} \code{List} object, which is used to specify options when estimating these models, to utilize the expectation maximization algorithm for optimization. As previously mentioned and described in the package documentation, the user can also set \code{method="MLE"} if they wish to employ maximum likelihood estimation. Additionally, we can specify whether we want the mean and variance to change according to the regime by setting \code{msmu=TRUE} and \code{msvar=TRUE}, respectively. Setting either of these to false would result in a model where that parameter is constant across regimes. The option \code{use\_diff\_init=30} is used to estimate the model thirty times with different initial values each time. The model with the highest log-likelihood value is kept as the main output, but the output \code{List} of these functions provides results for all iterations under the output \code{List} \code{trace}. The output \code{List} also contains various elements, all of which are described in the package documentation. Importantly, the outputs are \code{S3} objects, and the \code{print()} and \code{summary()} methods are provided for each, as can be seen in the example code above. Specifically, we see that the \code{summary()} method can be used to display the parameter estimates, which in this controlled setting can be compared to the true values, along with other characteristics such as the log-likelihood, AIC, BIC, and quantiles of the residuals.

\begin{figure}[!tbh]
\centering
\includegraphics{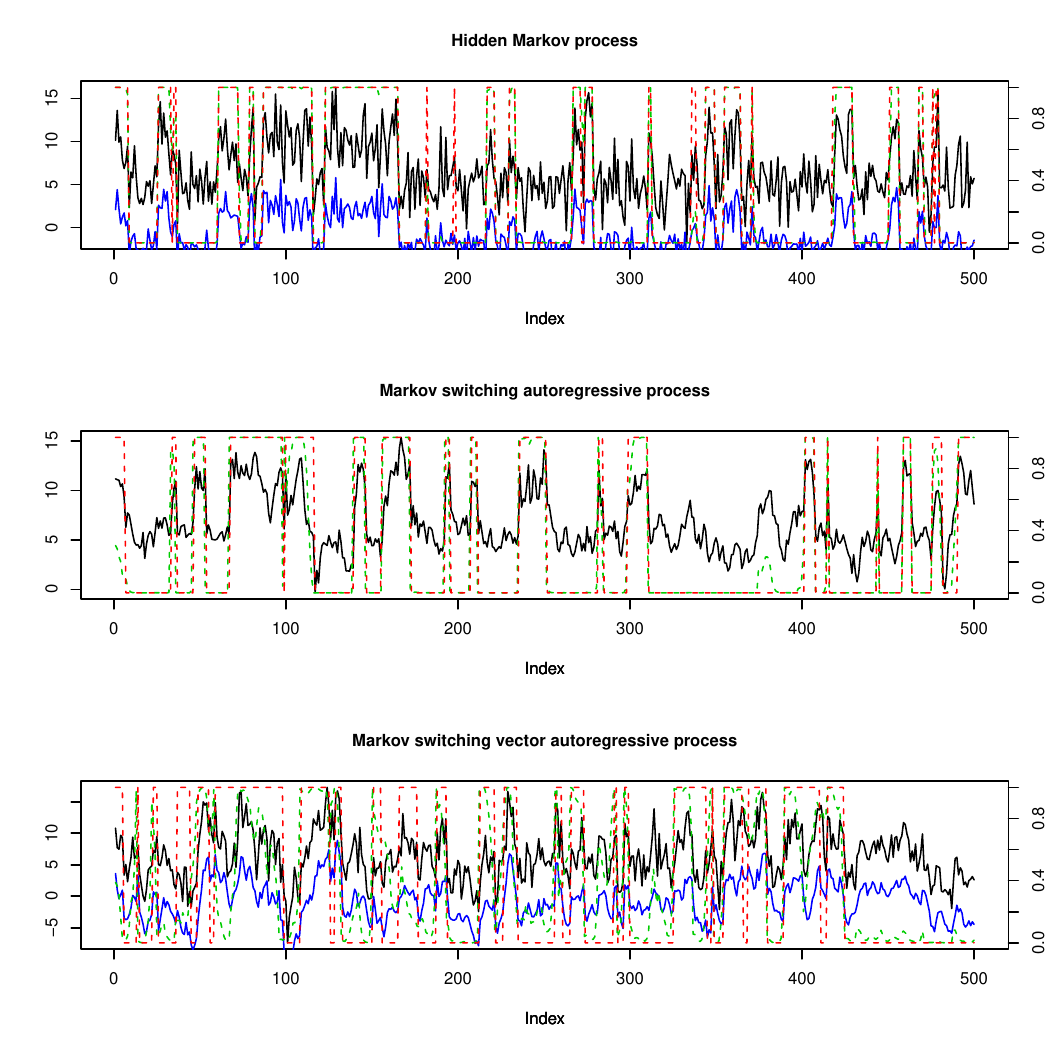}
\caption{\label{fig:estim} Simulated processes (black and blue), true regime states (red - dashed), and model estimated smoothed probabilities (green - dashed).}
\end{figure}

The Figure \ref{fig:estim} plots the simulated processes in black and blue, the true regime states $S_{t}$, which were also an output of the simulation functions, in red (dashed), and the model-estimated smoothed probabilities in green (dashed). From this, we can see that the detection of changes in regime is captured quite well in the estimation process. For some periods, there is, relatively speaking, a bit more difficulty with the Markov switching VAR model, but increasing sample sizes or considering more initial values could likely help in this regard.

% ----------------------------------------- %
% ----- Monte Carlo Likelihood Ratio Test
% ----------------------------------------- %
\subsection{Hypothesis testing}
In this section, we describe the hypothesis testing functions available in the \pkg{MSTest} package. The \pkg{MSTest} package has been designed with ease of use in mind and hence, in most cases, only requires the user to provide the variable $y_t$ (and $Z_t$ when relevant) and specify the number of regimes to test for.

\begin{table}[!tbh]
\centering
\begin{tabular}{p{0.2\textwidth}  p{0.7\textwidth} }
    \hline
    Function & Description \\
    \hline
    \code{LMCLRTest} & Local Monte Carlo Likelihood Ratio test proposed in \cite{rodrondufour_mcmstest}\\
    
    \code{MMCLRTest} & Maximized Monte Carlo Likelihood Ratio test proposed in \cite{rodrondufour_mcmstest}\\
    
    \code{DLMCTest}  &  \cite{dufourluger17} moment-based test for Markov switching autoregressive models. Extended to more general ARMA models and models with explanatory variables.  \\
    
    \code{DLMMCTest}  &  \cite{dufourluger17} Maximized Monte Carlo (MMC) version of the moment-based test. This function allows the user to set some explanatory variables as nuisance parameters. \\

    \code{CHPTest}  &   \cite{chp14} optimal test for Markov switching parameters.   \\

    %\code{CHPMCtest}  &   Monte Carlo Test version of \cite{chp14} optimal test for Markov switching parameters.   \\
    
    \code{HLRTest}	&   \cite{hansen92} likelihood ratio test. Uses empirical process theory to estimate asymtptotic distribution of bound for LR test. \\

    \hline
    \end{tabular}
\caption{\label{tab:testing} Available tests in the \proglang{R} package \pkg{MSTest}}
\end{table}

Table \ref{tab:testing} shows all the functions currently available in the \pkg{MSTest} package and briefly describes them. All functions return the test statistic, critical values of the test statistic distribution when available, the p-value, and parameter estimates under the null and alternative hypothesis depending on which were used for testing. It is important to note that, even though they are called critical values, the values returned by \code{HLRTest()} are the critical values of the process $Q$ discussed in \cite{hansen92}, which provide a bound for the LR but are not the critical values of the LR test for Markov switching. Likewise, the values returned by \code{LMCLRTest()}, \code{MMCLRTest()}, \code{DLMCTest()}, and \code{DLMMCTest()}, described as critical values, are in fact the percentiles of the simulated null distribution and hence we are using the term critical values loosely here. 

% ----------------------------------------- %
% ----- Monte Carlo Likelihood Ratio Test
% ----------------------------------------- %
\subsubsection{Monte Carlo likelihood ratio test}

We begin by discussing the Local Monte Carlo Likelihood Ratio test. As indicated in Table \ref{tab:testing}, this test can be implemented in the \pkg{MSTest} package using the function \code{LMCLRTest()}. Since this test requires estimating both the restricted and unrestricted models to simulate the null distribution, the options we would typically pass into the estimation functions using a \code{List} can be passed into the options of this testing procedure using \code{mdl\_h0\_control} and \code{mdl\_h1\_control}. The same applies to the Maximized Monte Carlo Likelihood Ratio test, which we describe next. Specifically, we can set options such as whether the mean or variance switches according to the regime, the number of initial values to use when estimating the models, and other related settings. However, the \code{LMCLRTest()} function allows users to specify a different number of initial values for estimating models during null distribution simulation through the \code{use\_diff\_init\_sim} option. By default, this is set to match the value used to estimate the model with observed data, as specified in the \code{mdl\_h0\_control} and \code{mdl\_h1\_control} \code{List}s, which is generally recommended.

Apart from the options that can be set, the \code{LMCLRTest()} function also requires specifying the number of lags, \code{p}, the number of regimes under the null hypothesis, \code{k0}, and the number of regimes under the alternative hypothesis, \code{k1}, to be compared. Below, we provide an example where we apply the LMC-LRT procedure to the Markov switching autoregressive model with $p=1$ lag, that we simulated above. In this example, we test the null hypothesis $H_0: M=1$ against the alternative $H_1: M=2$. Given that the simulated data is a Markov switching model with $M=2$ regimes, we would expect the null hypothesis to be rejected, which is indeed the result obtained.

\begin{CodeChunk}
\begin{CodeInput}
R> lmc_control = list(N = 99,
+                     mdl_h0_control = list(const  = TRUE, 
+                                           getSE  = FALSE),
+                     mdl_h1_control = list(msmu   = TRUE, 
+                                           msvar  = TRUE,
+                                           getSE  = FALSE,
+                                           method = "EM",
+                                           use_diff_init = 5))
R> 
R> lmclrt <- LMCLRTest(simu_msar[["y"]], p = 1, k0 = 1 , k1 = 2, control = lmc_control)
R> summary(lmclrt)

Restricted Model
        coef
mu    6.8649
phi_1 0.8330
sig   2.9119

log-likelihood =  -973.7215
AIC =  1953.443
BIC =  1966.081

Unrestricted Model
           coef
mu_1   5.246300
mu_2  10.754000
phi_1  0.788720
sig_1  0.937860
sig_2  1.935100
p_11   0.949260
p_12   0.050742
p_21   0.119250
p_22   0.880750

log-likelihood =  -863.7258
AIC =  1745.452
BIC =  1783.365

Rodriguez-Rondon & Dufour (2024) Local Monte Carlo Likelihood Ratio Test
         LRT_0  0.90%  0.95%  0.99% p-value
LMC_LRT 219.99 8.3114 9.1987 10.922    0.01
\end{CodeInput}
\end{CodeChunk}

As previously mentioned, the Local Monte Carlo Likelihood Ratio test function, \code{LMCLRTest()}, can be used to replicate the parametric bootstrap approach discussed in \cite{quzhuo2021likelihood} and \cite{kasshi2018}. To do this, we start by setting \code{mdl\_h1\_control = list(method = "MLE")}. In those studies, MLE is used, and using MLE in \pkg{MSTest} also allows us to constrain the parameter space. To set parameter constraints, we define the vectors \code{mle\_theta\_low} and \code{mle\_theta\_upp} in the \code{mdl\_h1\_control} \code{List}. In addition to constraining the parameter space for the transition probabilities, \cite{kasshi2018} also impose restrictions on the variance, which can be implemented here. Additionally, we may wish to increase the value of \code{lmc\_control = list(N)} to improve the approximation of the asymptotic critical values. For example, \cite{quzhuo2021likelihood} set $N=199$, while \cite{kasshi2018} set $N=299$ for similar purposes. Ideally, for a bootstrap procedure, this value should be higher. However, given the computational demands of estimating Markov switching models, using lower values can be reasonable.

The Maximized Monte Carlo Likelihood Ratio test can be conducted using the \code{MMCLRTest()} function. This function shares the same options as the LLMC-LRT procedure for estimating both the restricted and unrestricted models. Similarly, users must specify the number of lags, \code{p}, the number of regimes under the null hypothesis, \code{k0}, and the number of regimes under the alternative hypothesis, \code{k1}, for comparison. However, \code{MMCLRTest()} also offers additional settings related to the testing procedure, as this method is more intricate. For instance, users can set the fixed constant $c$, used to define the consistent set over which to search, by setting \code{eps}. To use the set $C_{T}^{\ast} = C_{T}^{CI} \cup C_{T}^{\epsilon}$, as described earlier, \code{CI\_union=TRUE} should be specified. To use only $C_{T}^{CI}$, one can set \code{eps=0} and \code{CI\_union=TRUE}. Additionally, users can select the numerical optimization algorithm for this search via \code{type}. By default, the algorithm stops when a p-value of $1$ is reached, as this is the highest possible p-value. Alternatively, users can choose to stop when the test fails to reject, that is, upon reaching any value above the test level $\alpha$, by using \code{threshold\_stop}.

Below, we provide an example where the MMC-LRT procedure is used to test a linear autoregressive model with only $M=1$ regime. Here, we set \code{eps=0.3} and \code{CI\_union=TRUE} to use the consistent set $C_{T}^{\ast }$ and set \code{type="pso"} to employ the particle swarm algorithm (see \cite{psopack}). Other available optimization algorithms include: Simulated Annealing using the ``GenSA'' package introduced by \cite{xiaetal2013}, Particle Swarm algorithm using the ``PSO'' package introduced by\cite{zametal2013} and Genetic algorithm using the ``GA'' introduced by \cite{scrucca2013}. Importantly, we also set \code{workers=8}, which allows us to use a parallel version of the test to improve computational efficiency. This option is also available for the \code{LMCLRTest()} function. Specifically, the null distribution is simulated using $8$ different workers in this case. Note that, as shown in the example, the user must register a parallel pool and close the cluster in order to make use of this functionality.

\begin{CodeChunk}
\begin{CodeInput}
R> mmc_control = list(N = 99,
+                     eps = 0.3,
+                     threshold_stop = 0.05 + 1e-6, 
+                     type = "pso",
+                     workers  = 8,
+                     CI_union = FALSE,
+                     mdl_h0_control = list(const  = TRUE, 
+                                           getSE  = FALSE),
+                     mdl_h1_control = list(msmu   = TRUE, 
+                                           msvar  = TRUE,
+                                           getSE  = FALSE,
+                                           method = "EM"),
+                     maxit  = 100)
R> 
R> doParallel::registerDoParallel(mmc_control[["workers"]])
R> mmclrt <- MMCLRTest(simu_ar[["y"]], p = 1, k0 = 1 , k1 = 2, control = mmc_control)
S=13, K=3, p=0.2135, w0=0.7213, w1=0.7213, c.p=1.193, c.g=1.193
v.max=NA, d=1.039, vectorize=FALSE, hybrid=off
It 1: fitness=-0.95
Converged
R> summary(mmclrt)

Restricted Model
         coef
mu    4.86110
phi_1 0.71919
sig   1.04530

log-likelihood =  -718.0963
AIC =  1442.193
BIC =  1454.83

Unrestricted Model
            coef
mu_1  7.6298e+00
mu_2  4.8671e+00
phi_1 7.1919e-01
sig_1 7.2678e-01
sig_2 1.0411e+00
p_11  4.3003e-13
p_12  1.0000e+00
p_21  4.4665e-07
p_22  1.0000e+00

log-likelihood =  -718.095
AIC =  1454.19
BIC =  1492.103

Rodriguez-Rondon & Dufour (2024) Maximized Monte Carlo Likelihood Ratio Test
            LRT_0 p-value
MMC_LRT 0.0024992    0.95
R> 
R> doParallel::stopImplicitCluster()
\end{CodeInput}
\end{CodeChunk}

As expected, we fail to reject the null hypothesis of no Markov switching (i.e., $M=1$) in this case since the true DGP is one of a linear autoregressive process. Given that we set \code{threshold\_stop = 0.05 + 1e-6}, the algorithm converged quickly. We could continue searching for the parameter values under the null that give the maximum p-value, but for the purpose of exposition, we set the \code{threshold} parameter so that we stop searching once the test fails to reject the null hypothesis.

% ----------------------------------------- %
% ----- DL Moment-based tests
% ----------------------------------------- %
\subsubsection{Moment-based tests}

The Monte Carlo moment-based test proposed by \cite{dufourluger17} is invoked using \code{DLMCTest()} (local version) and \code{DLMMCTest()} (maximized version). Like all other functions, these require $y_{t}$ and $p$, the number of autoregressive lags, to be specified by the user. The parameter $N$ sets the number of Monte Carlo replications. The default value for $N$ is $99$, as \cite{dufour2004} shows that having more than $100$ replications (i.e., including the value obtained from the observed series) has a minimal effect on power. This test also involves approximating the distribution of the p-value for each statistic through a separate round of simulation. The number of replications used in this approximation is determined by $N2$, which is set to $10,000$ by default. For the moment-based local Monte Carlo test, these are the only required and available options.

Below, we provide an example of this test using the previously simulated Markov switching autoregressive data. Results are obtained very quickly. The output of the \code{summary()} function presents the nuisance parameter value (in this case, $\phi_1$), the test statistic for each of the four moments, the combined test statistic under $F(e)$, the critical values, and the Monte Carlo p-values. Here, we observe a clear rejection of the null hypothesis, which is expected since the process used is indeed a Markov switching process with $M=2$ regimes.

\begin{CodeChunk}
\begin{CodeInput}
R> lmc_control = list(N = 99,
+                     simdist_N = 10000,
+                     getSE = TRUE)
R> 
R> lmcmoment <- DLMCTest(simu_msar[["y"]], p = 1, control = lmc_control)
R> summary(lmcmoment)

Restricted Model
        coef     s.e.
mu    6.8649 0.457460
phi_1 0.8330 0.024688
sig   2.9119 0.185100

log-likelihood =  -973.7215
AIC =  1953.443
BIC =  1966.081

Dufour & Luger (2017) Moment-Based Local Monte Carlo Test
         phi_1 M(e)   V(e)    S(e)   K(e) F(e)   0.90%   0.95%   0.99% p-value
LMC_min  0.833 1.37 16.344 0.24838 3.3067    1 0.96814 0.98551 0.99557    0.01
LMC_prod 0.833 1.37 16.344 0.24838 3.3067    1 0.99915 0.99969 0.99981    0.01
\end{CodeInput}
\end{CodeChunk}

The computational efficiency of this procedure extends well to the Moment-based maximized Monte Carlo test proposed by \cite{dufourluger17}, which again can be performed very quickly using the \pkg{MSTest} package. Here, more options are available regarding the optimization over the nuisance parameter space. Here, \code{optim\_type} is used to determine the numerical optimization algorithm to be used. As with the MMC-LRT procedure, we can use \code{eps} and \code{CI\_union} to define the consistent set over which to maximize the p-value. 

The computational efficiency of this procedure extends well to the moment-based maximized Monte Carlo test proposed by \cite{dufourluger17}, which can also be performed very quickly using the \pkg{MSTest} package, despite the maximization over the nuisance parameter space. In this case, like with the MMC-LRT procedure, more options are available regarding the optimization over this space. Specifically, \code{optim\_type} is used to determine the numerical optimization algorithm to be employed. Similar to the MMC-LRT procedure, we can utilize \code{eps} and \code{CI\_union} to define the consistent set over which to maximize the p-value. In fact, many of the same options available for the \code{MMCLRTest()} and available here also. 

\begin{CodeChunk}
\begin{CodeInput}
R> mmc_control <- list(N = 99,
+                      getSE = TRUE,
+                      eps = 0, 
+                      CI_union = TRUE,
+                      optim_type = "GenSA",
+                      threshold_stop = 0.05 + 1e-6, 
+                      maxit = 100)
R> 
R> mmcmoment <- DLMMCTest(simu_msar[["y"]], p = 1, control = mmc_control)
Stop. Nb function call=100 max function call=100.
Emini is: -0.01
xmini are:
0.8330039843 
Totally it used 0.001308 secs
No. of function call is: 100
Stop. Nb function call=100 max function call=100.
Emini is: -0.01
xmini are:
0.8330039843 
Totally it used 0.001065 secs
No. of function call is: 100
R> summary(mmcmoment)

Restricted Model
        coef     s.e.
mu    6.8649 0.457460
phi_1 0.8330 0.024688
sig   2.9119 0.185100

log-likelihood =  -973.7215
AIC =  1953.443
BIC =  1966.081

Dufour & Luger (2017) Moment-Based Maximized Monte Carlo Test
         M(e)   V(e)    S(e)   K(e) F(e) p-value
MMC_min  1.37 16.344 0.24838 3.3067    1    0.01
MMC_prod 1.37 16.344 0.24838 3.3067    1    0.01
\end{CodeInput}
\end{CodeChunk}

Above, we provide an example of this function, where we again utilize the Markov switching autoregressive process that was previously simulated. We set the \code{threshold} parameter to be $0.05 + 1e-6$ so that we stop searching if the test fails to reject the null hypothesis, which should not be the case. This time, we also set \code{eps=0} so that the search is performed over the confidence interval, as in \cite{dufourluger17}. Here, we reject the null hypothesis of a linear model once more, a result that is consistent with the tests used so far.

% ----------------------------------------- %
% ----- CHP Parameter stability test
% ----------------------------------------- %
\subsubsection{Parameter stability test}
%\paragraph{CHPtest()}
The test proposed by \cite{chp14} can be employed by using \textbf{CHPTest()}, where again $y_{t}$ and $p$ must be provided by the user. In this case, $N$ is the number of bootstraps and is set to $3000$ by default, as in their work, and $\rho$ is used to set the bound for one of the nuisance parameters. This value is set to $0.7$ by default, as in their work also. By setting $\rho = 0.7$, the grid search occurs over the parameter space $\rho \in [-0.7, 0.7]$.

\begin{CodeChunk}
\begin{CodeInput}
R> chp_control = list(N = 1000, 
+                     rho_b = 0.7, 
+                     msvar = TRUE)
R> 
R> pstabilitytest <- CHPTest(simu_ar[["y"]], p = 1, control = chp_control)
R> summary(pstabilitytest)

Restricted Model
         coef     s.e.
mu    4.86110 0.162990
phi_1 0.71919 0.031258
sig   1.04530 0.066442

log-likelihood =  -718.0963
AIC =  1442.193
BIC =  1454.83

Carrasco, Hu, & Ploberger (2014) Parameter Stability Test 

- Switch in Mean and Variance
      test-stat  0.90%  0.95%  0.99% p-value
supTS   0.65892 2.6918 3.2964 5.6429   0.676
expTS   0.65116 1.2877 1.6056 3.0695   0.533
\end{CodeInput}
\end{CodeChunk}

Above, we provide an example of this test using the linear autoregressive process simulated previously again. In this case, we also set \code{msvar=TRUE} to consider a test where changes in the variance are also considered. The \code{summary()} function provides estimates of the restricted model, as this is the only model needed to perform the test, and displays results from the supTS and expTS versions of the test. As expected, we also find that the test fails to reject the null hypothesis of a linear model.

% ----------------------------------------- %
% ----- Hansen 1992 Likelihood Ratio Test
% ----------------------------------------- %
\subsubsection{Stochastic likelihood ratio test}

Finally, in the \pkg{MSTest} package, the test proposed by \cite{hansen92} can be invoked using the command \textbf{HLRTest()}. As before, this function requires the user to provide $y_{t}$, the variable of interest. This test is designed to assess the null hypothesis of linearity in autoregressive models, so a value for the number of autoregressive components $p$ must again be provided. The test performs a grid search over the nuisance parameters, meaning most of the available options are related to configuring this grid. Specifically, users can set the \code{gridsize}, as well as the starting point and step size for both the mean and variance grids. These values pertain to the unrestricted model. While keeping these values fixed, the test procedure optimizes the values of the restricted model; thus, the user can utilize \code{theta\_null\_low} and \code{theta\_null\_upp} to define the optimization space. It is important to note that while setting \code{msvar = TRUE} is possible, it will require more computation time as the function must optimize over a larger parameter space.

\begin{CodeChunk}
\begin{CodeInput}
R> hlrt_control  <- list(msvar          = TRUE,
+                        gridsize       = 20,
+                        mugrid_from    = 0,
+                        mugrid_by      = 1,
+                        siggrid_from   = 0.5,
+                        siggrid_by     = 0.1,
+                        theta_null_low = c(0,-0.99,0.01),
+                        theta_null_upp = c(20,0.99,20))
R> # Perform Hansen (1992) likelihood ratio test
R> hlrt <- HLRTest(simu_msar[["y"]], p = 1, control = hlrt_control)
R> summary(hlrt)

Restricted Model
        coef     s.e.
mu    6.8649 0.457460
phi_1 0.8330 0.024688
sig   2.9119 0.185100

log-likelihood =  -973.7215
AIC =  1953.443
BIC =  1966.081

Hansen (1992) Likelihood Ratio Bound Test -  Switch in Mean and Variance
      test-stat 0.90 % 0.95 % 0.99 % p-value
M = 0    9.2127 2.7604 3.0515 3.7031       0
M = 1    9.2127 2.8116 3.0626 3.6341       0
M = 2    9.2127 2.8518 3.1555 3.6657       0
M = 3    9.2127 2.9257 3.2456 3.7498       0
M = 4    9.2127 2.9979 3.2328 3.8840       0
\end{CodeInput}
\end{CodeChunk}

Above, we provide an example of this test using the previously simulated Markov switching autoregressive process. In this case, we also set \code{msvar=TRUE} to consider a test where changes in variance are enabled. The \code{summary()} function provides estimates of the restricted model, as this is the only model needed to perform the test in this instance. This output also displays results from the different types of standardizations mentioned in \cite{hansen96}. In all cases, we find that the test rejects the null hypothesis of a linear model, yielding a p-value that is very small, approximately $0$.
 
% ----------------------------------------- %
% ----- Other functionalities
% ----------------------------------------- %
%\subsection{Other functionalities}

%% -- Empirical Example -------------------------------------------------------

%% - Virtually all JSS manuscripts list source code along with the generated
%%   output. The style files provide dedicated environments for this.
%% - In R, the environments {Sinput} and {Soutput} - as produced by Sweave() or
%%   or knitr using the render_sweave() hook - are used (without the need to
%%   load Sweave.sty).
%% - Equivalently, {CodeInput} and {CodeOutput} can be used.
%% - The code input should use "the usual" command prompt in the respective
%%   software system.
%% - For R code, the prompt "R> " should be used with "+  " as the
%%   continuation prompt.
%% - Comments within the code chunks should be avoided - these should be made
%%   within the regular LaTeX text.

\section{Empirical example} \label{sec:empiricalex}
In this section, we apply the full suite of tests available in \pkg{MSTest} to three empirical data sets provided in the package, specifically to three samples of U.S. GNP growth rates covering distinct periods of interest. According to \cite{kimnel1999}, a model with two regimes should capture the structural decline in business cycle volatility that began in the mid-1980s, a phenomenon known as the \textit{Great Moderation}. Additionally, as discussed by \cite{hamilton89}, the presence of business cycle fluctuations support a two-regime model to to reflect shifts in the growth rate of U.S. GNP. These characteristics make U.S. GNP growth an ideal empirical example for the tests provided by \pkg{MSTest}. If the model accurately captures these structural shifts, by allowing both the mean and variance to vary by regime we should expect the tests to at least reject the null hypothesis of one regime (i.e., a linear model) for the second and third sample, which covers both the high and low-volatility periods and various recessionary and expansionary periods. 

We conduct the tests using two model specifications. In the first, only the mean changes by regime, as initially suggested by \cite{hamilton89}. IN the second, and an alternative specification where both the mean and variance vary by regime is considered. Table \ref{tab:empiricaltest} presents the test results for all three samples across these two specifications. The first panel shows results when only the mean changes by regime. Here, we observe that the tests fail to reject the null hypothesis of a linear model for the first sample but reject it for the third, extended sample. For the second sample, covering 1951Q2 to 2010Q4, the supTS, expTS, and H-LRT tests fail to reject the null hypothesis of a linear model, while the LMC-LRT and MMC-LRT procedures reject it. Simulation evidence from \cite{rodrondufour_mcmstest} suggests that in settings where only the mean changes, the LMC-LRT and MMC-LRT tests generally have higher power and may provide more reliable results. The second panel shows results when both the mean and variance vary by regime. Here, all tests are consistent. That is, they suggest that there is insufficient evidence to reject the null hypothesis of a linear model in the first sample, but find sufficient evidence against this null hypothesis in the two larger samples. 

While it would be valuable to further investigate the potential presence of a third regime in the second and third samples, this analysis is thoroughly conducted in \cite{rodrondufour_mcmstest}. In their work, the authors also consider controlling for the Great Moderation and COVID periods by treating these as known structural breaks in the mean, offering a more comprehensive analysis. As they employ the \pkg{MSTest} package for these tests, we direct interested readers to their paper for these results.

\begin{table}[!tbh]
\centering
\begin{tabular}{p{0.2\textwidth}  p{0.10\textwidth} p{0.10\textwidth} p{0.10\textwidth} p{0.10\textwidth} p{0.10\textwidth} p{0.1\textwidth}}
    \hline
    \multirow{ 2}{*}{Test procedure} & \multicolumn{2}{c}{1951Q2-1984Q4} & \multicolumn{2}{c}{1951Q2-2010Q4} & \multicolumn{2}{c}{1951Q2-2024Q2} \\
    & test-stat & p-value & test-stat & p-value & test-stat & p-value \\
    \hline
    \multicolumn{7}{c}{Panel A: Change in mean}\\
    \hline
    LMC-LRT    &  4.83     &  0.30     & 15.72      &  0.01     & 99.64      &  0.01  \\
    MMC-LRT    &  4.83     &  0.46     & 20.82      &  0.02     & 99.64      &  0.01  \\
    supTS      &  0.00     &  0.28     &  0.31      &  0.15     & 0.10       &  0.21  \\
    expTS      &  0.63     &  0.18     &  0.33      &  0.49     &  0.63      &  0.18  \\
    H-LRT      &  1.91     &  0.46     &  1.75      &  0.60     & 1.79       &  0.39  \\
    \hline
    \multicolumn{7}{c}{Panel B: Change in mean \& variance}\\
    \hline
    LMC-LRT              & 9.14      & 0.18      & 50.19      & 0.01      & 195.04      & 0.01  \\
    MMC-LRT              & 9.14      & 0.25      & 50.19      & 0.01      & 195.04      & 0.01  \\
    LMC$_\text{min}$     & 0.82      & 0.63      &  1.00      & 0.01      &   1.00      & 0.01  \\
    LMC$_\text{prod}$    & 0.96      & 0.68      &  1.00      & 0.01      &   1.00      & 0.01  \\
    MMC$_\text{min}$     & 0.37      & 1.00      &  1.00      & 0.01      &   1.00      & 0.01  \\
    MMC$_\text{prod}$    & 0.79      & 0.99      &  1.00      & 0.02      &   1.00      & 0.01  \\
    supTS                & 1.67      & 0.19      & 15.18      & 0.00      &  34.21      & 0.00  \\
    expTS                & 0.85      & 0.22      & 330.75    &  0.00      & 5.05x$10^9$      & 0.00  \\
    H-LRT                & 1.61      & 1.00      & 5.07      &  0.00      & 10.75      &  0.00  \\

    \hline
    \end{tabular}
\caption{\label{tab:empiricaltest} Hypothesis test results with data sets available in \pkg{MSTest}.}
\end{table}

%% -- Summary/conclusions/discussion -------------------------------------------

\section{Conclusion} \label{sec:conclusion}
The importance of testing the number of regimes in Markov switching models has led to numerous contributions, each addressing the statistical and computational challenges inherent in this problem. Notable works include \cite{hansen92}, \cite{chp14}, and \cite{dufourluger17}, which focus on testing the null hypothesis of a single regime (i.e., a linear model) versus the alternative of two regimes. More recently, \cite{rodrondufour_mcmstest} introduced a set of Monte Carlo test procedures for testing a null hypothesis of $M_0$ regimes against an alternative of $M_0+m$ regimes, applicable when both $M_0 \geq 1$ and $m \geq 1$. The \proglang{R} package \pkg{MSTest} makes these test procedures available, implementing the methods from these four studies. This paper reviews these procedures and explains how \pkg{MSTest} can be used to perform these tests, as well as other package features such as simulation and model estimation. The goal of \pkg{MSTest} is to provide researchers with a user-friendly tool for conducting these tests, facilitating inference, and determining the appropriate model specifications for their data.
%% -- Optional special unnumbered sections -------------------------------------

\section*{Computational details}

The results in this paper were obtained using
\proglang{R}~4.4.0 \cite{r2024} with the packages
\pkg{MSTest}~0.1.3 \cite{MSTest2024}, \pkg{Rcpp}~1.0.13 \cite{eddetal2018}, \pkg{RcppArmadillo}~14.0.2.1 \cite{eddsan2024}, \pkg{GenSA}~1.1.14.1 \cite{xiaetal2013}, \pkg{pso}~1.0.4 \cite{psopack}, \pkg{foreach}~1.5.2 \cite{foreachpack}, and \pkg{doParallel}~1.0.17 \cite{doparpack}. Computations were performed on Apple macOS Sonoma Version 14.2.1 aarch64-apple-darwin20. Code for the computations is available in the R script article.R, available in the GitHub repository at \url{https://github.com/roga11/MSTest/tree/main/inst/examples}. \proglang{R} itself
and all packages used are available from the Comprehensive
\proglang{R} Archive Network (CRAN) at
\url{https://CRAN.R-project.org/}.

\section*{Acknowledgments}

This work was supported by the Fonds de recherche sur la société et la culture Doctoral Research Scholarships (B2Z).

%% -- Bibliography -------------------------------------------------------------
%% - References need to be provided in a .bib BibTeX database.
%% - All references should be made with \cite, \citet, \citep, \citealp etc.
%%   (and never hard-coded). See the FAQ for details.
%% - JSS-specific markup (\proglang, \pkg, \code) should be used in the .bib.
%% - Titles in the .bib should be in title case.
%% - DOIs should be included where available.

%\printbibliography

\bibliographystyle{apalike}
\bibliography{refs.bib}

%% -----------------------------------------------------------------------------

\end{document}